\documentclass[english,aps,superscriptaddress]{revtex4}
\usepackage[T1]{fontenc}
\usepackage[latin9]{inputenc}
\usepackage{mathrsfs}
\usepackage{amsmath}
\usepackage{amssymb}
\usepackage{esint}

\makeatletter

\providecommand{\tabularnewline}{\\}

\@ifundefined{textcolor}{}
{%
 \definecolor{BLACK}{gray}{0}
 \definecolor{WHITE}{gray}{1}
 \definecolor{RED}{rgb}{1,0,0}
 \definecolor{GREEN}{rgb}{0,1,0}
 \definecolor{BLUE}{rgb}{0,0,1}
 \definecolor{CYAN}{cmyk}{1,0,0,0}
 \definecolor{MAGENTA}{cmyk}{0,1,0,0}
 \definecolor{YELLOW}{cmyk}{0,0,1,0}
}

\usepackage{babel}

\makeatother

\usepackage{babel}
\begin{document}

\title{A quark coalescence model for polarized vector mesons and baryons}

\author{Yang-guang Yang}

\affiliation{Department of Modern Physics, University of Science and Technology
of China, Hefei, Anhui 230026, China}

\author{Ren-hong Fang}

\affiliation{Key Laboratory of Quark and Lepton Physics (MOE) and Institute of
Particle Physics, Central China Normal University, Wuhan, 430079,
China}

\author{Qun Wang}

\affiliation{Department of Modern Physics, University of Science and Technology
of China, Hefei, Anhui 230026, China}

\author{Xin-nian Wang}

\affiliation{Key Laboratory of Quark and Lepton Physics (MOE) and Institute of
Particle Physics, Central China Normal University, Wuhan, 430079,
China}

\affiliation{Nuclear Science Division, MS 70R0319, Lawrence Berkeley National
Laboratory, Berkeley, California 94720, USA}
\begin{abstract}
A non-relativistic quark coalescence model is formulated for polarized
vector mesons and baryons of spin-1/2 octet and spin-3/2 decuplet.
With the spin density matrix, one can compute in a uniform way the
polarizations of vector mesons and baryons from those of quarks and
antiquarks with explicit momentum dependence. The results are compared
to that obtained from kinetic and statistical models for hadrons. 
\end{abstract}
\maketitle

\section{Introduction}

The system in non-central heavy-ion collisions at high energies have
large orbital angular momenta \cite{Liang:2004ph,Liang:2004xn,Voloshin:2004ha,Betz2007,Becattini:2007sr,Gao2008}
(see Ref. \cite{Wang:2017jpl} for a recent review). Some of the angular
momenta are transferred to the hot and dense matter and lead to polarization
of quarks along the direction of the orbital angular momentum or the
event plane. This type of polarization is called the global polarization.
Recently the STAR collaboration has measured the global polarization
of $\Lambda$ and $\bar{\Lambda}$ hyperons in the beam energy scan
program \cite{STAR:2017ckg,Abelev:2007zk}. At all energies below
62.4 GeV, non-vanishing values of the global polarization for $\Lambda$
and $\bar{\Lambda}$ have been measured \cite{STAR:2017ckg}.

Strong magnetic fields are also produced in non-central heavy ion
collisions \cite{Kharzeev:2007jp,Skokov:2009qp,Voronyuk:2011jd,Deng:2012pc,Bloczynski:2012en,McLerran:2013hla,Gursoy:2014aka,Roy:2015coa,Tuchin:2014iua,Li:2016tel}.
The magnetic fields are on the average in the same direction as the
orbital angular momenta. But such strong magnetic fields decay very
quickly after two nuclei pass through each other at high energies.
So they are expected to have effects only in the partonic phase during
the early stage of the evolution \cite{Vilenkin:1980fu,Kharzeev:2007jp,Fukushima:2008xe,Son:2009tf,Son:2012bg,Son:2012wh,Stephanov:2012ki,Gao:2012ix}.
The strong magnetic fields can also polarize quarks through their
magnetic moments \cite{Fang:2016vpj,Mueller:2017arw}.

How these polarized quarks form and relate to polarized hadrons is
an important question to understand the polarization phenomenon in
the final state of hadrons. Most previous calculations of the global
polarizations of the $\Lambda$ and $\bar{\Lambda}$ hyperons \cite{Becattini:2013vja,Becattini:2016gvu,Karpenko:2016jyx,Li:2017slc,Xie:2016fjj,Xie:2017upb}
are based on statistical-hydro models \cite{weert:1982,zubarev:1979,Becattini:2009wh,Becattini:2012tc,Becattini:2013fla,Becattini:2015nva,Hayata:2015lga}.
In these calculations the polarizations of hadrons are assumed to
arise from the thermal (Fermi-Dirac) distribution with spin degrees
of freedom due to the vorticity fields which are obtained by hydrodynamic
\cite{Csernai:2013bqa,Csernai:2014ywa,Pang:2016igs} or transport
models \cite{Jiang:2016woz,Li:2017slc,Sun:2017xhx}. In an early paper
\cite{Liang:2004ph} by Liang and one of us, an estimate of the hyperon
polarization from polarized quarks was made in a quark recombination
model. The same quark recombination model was used to compute the
spin density matrix elements of vector mesons \cite{Liang:2004xn}.
But momentum dependence was not included in such a quark recombination
model.

In this paper we will formulate a non-relativistic quark coalescence
model with explicit momentum dependence based on the spin density
matrix, with which we can compute the polarizations of vector mesons
and baryons of spin-1/2 octet and spin-3/2 decuplet through coalescence
of polarized quarks and anti-quarks in a systematic way. The conventional
quark coalescence models \cite{Greco:2003xt,Fries:2003kq} or recombination
models \cite{Xie:1988wi,Wang:2012cw} do not include the spin degrees
of freedom. We focus on vector mesons and baryons of the spin-1/2
octet and spin-3/2 decuplet. The generalization to other meson and
baryon multiplets are straightforward. With this coalescence model
we can compute the polarizations of vector mesons and baryons as functions
of their momenta from the quark polarizations obtained from other
approaches such as Wigner function approach \cite{Gao:2012ix,Chen:2012ca,Son:2012zy,Gao:2015zka,Fang:2016vpj,Hidaka:2016yjf,Mueller:2017lzw}.
The polarizations include those induced by the vorticity and by the
magnetic field.

The paper is organized as follows. In section \ref{sec:density-matrix},
we introduce the density matrix which we will use to compute a particle's
polarization. In Section \ref{sec:density-meson}, we formulate a
quark coalescence model for the spin density matrix of vector mesons
with explicit momentum dependence. In Section \ref{sec:baryon}, we
formulate the same model for the polarizations of baryons in the spin-1/2
octet and spin-3/2 decuplet. In Section \ref{sec:wigner} we show
how to compute the quark polarization from Wigner functions. In Section
\ref{sec:quark-polar} we discuss in some approximations the quark
and hadron polarizations from the vorticity and from the magnetic
field separately. We also give some features of the spin density matrix
element $\rho_{00}$ for vector mesons. In the last section we conclude
by giving a summary of the main results.

\section{Basics of spin density matrix}

\label{sec:density-matrix}Since we make use of the spin density matrix
throughout the paper to compute the particle polarization, in this
section we give some basics about the spin density matrix.

We consider an ensemble of particles with spin quantum number $S$.
The normalized spin states are labeled by $|\psi_{i}\rangle$ from
which the spin density operator can be defined 
\begin{equation}
\rho=\sum_{i}P_{i}|\psi_{i}\rangle\langle\psi_{i}|,\label{eq:density-operator}
\end{equation}
where $P_{i}$ is the probability of the spin state $|\psi_{i}\rangle$.
We list here the properties of the spin density operator \cite{Minnaert:1966}:
(a) $\rho$ is Hermitian, $\rho=\rho^{\dagger}$; (b) The trace of
$\rho$ is 1, $\text{Tr}\rho=\sum_{j}\left\langle \psi_{j}\right|\rho\left|\psi_{j}\right\rangle =\sum_{i}P_{i}=1$;
(c) $\rho$ is a positive-semidefinite operator, i.e., for any state
$|\phi\rangle$, we have $\langle\phi|\rho|\phi\rangle=\sum_{i}P_{i}\bigg|\langle\phi|\psi_{i}\rangle\bigg|^{2}\geqslant0$.

The dimension of the spin space for a spin-$S$ particle is $(2S+1)$,
so the spin density operator is a $(2S+1)\times(2S+1)$ matrix with
$2(2S+1)^{2}$ real parameters. The conditions that $\rho$ is Hermitian
and has unit trace reduce the number of real parameters to $4S(S+1)$.
The positive-semidefinite condition can also impose restrictions on
these real parameters.

If the particle is unpolarized, the spin density matrix $\rho$ has
a simple form 
\begin{equation}
\rho=\frac{1}{2S+1}\text{diag}\,(1,1,\cdots,1).\label{eq:you3}
\end{equation}
Any deviation from (\ref{eq:you3}) indicates some degree of spin
alignment or polarization.

Let us look at a few examples. The first example is the spin-1/2 particle.
The spin density matrix has three real parameters and can be written
as 
\begin{equation}
\rho=\frac{1}{2}(1+\overrightarrow{\mathscr{P}}\cdot\boldsymbol{\sigma}),
\end{equation}
where $\overrightarrow{\mathscr{P}}$ is a unit three-vector serving
as the polarization vector and $\boldsymbol{\sigma}=(\sigma_{1},\sigma_{2},\sigma_{3})$
are Pauli matrices. The polarization can be read out by 
\begin{equation}
\overrightarrow{\mathscr{P}}=\frac{\text{Tr}(\rho\boldsymbol{\sigma})}{\text{Tr}(\rho)}.
\end{equation}
The octet hyperon $\Lambda^{0}$ is a spin-1/2 particle whose polarization
can be measured by its weak decay, $\Lambda^{0}\rightarrow\mathrm{p}+\pi^{-}$
\cite{Bunce:1976yb,Abelev:2007zk}. In Section \ref{sec:baryon} we
will use the quark coalescence model to compute the polarization of
$\Lambda^{0}$ through its spin density matrix.

The second example is the spin-1 particle whose spin density matrix
has eight real parameters. We look at the vector meson $K^{*0}$ and
$\phi$ for illustration with following strong decay modes 
\begin{eqnarray}
K^{*0} & \rightarrow & K^{+}+\pi^{-},\ \ (\sim100\%),\nonumber \\
\phi & \rightarrow & K^{+}+K^{-},\ \ (\sim49\%),\label{eq:decay-vector}
\end{eqnarray}
where the fractions are the branching ratios. The polarization of
$K^{*0}$ and $\phi$ can be measured through these decays \cite{Abelev:2008ag,BedangadasMohantyfortheALICE:2017xgh}.
We note that parity is conserved in strong decays so the final states
must be in the p-wave with $L=1$. We take $K^{*0}$ for an example,
all discussions apply to $\phi$ equally. The decay amplitude can
be described by the $\mathscr{S}$ matrix 
\begin{equation}
\langle K^{+},\pi^{-}|\mathscr{S}|K^{*0};S_{z}\rangle=Y_{1,S_{z}}(\theta,\phi)
\end{equation}
where $S_{z}=-1,0,1$, $(\theta,\phi)$ denote the polar and azimuthal
angle of one decay daughter, and $Y_{1,S_{z}}(\theta,\phi)$ denote
the spherical harmonic functions of $L=1$. The angular distribution
of $K^{+}$ or $\pi^{-}$ for a specific initial spin state $\left|1,S_{z}\right\rangle $
of $K^{*0}$ is 
\begin{equation}
\frac{dN}{d\Omega}=\bigg|\langle K^{+},\pi^{-}|\mathscr{S}|K^{*0};S_{z}\rangle\bigg|^{2}=\left|Y_{1,S_{z}}(\theta,\phi)\right|^{2}.
\end{equation}
We now consider an ensemble of $K^{*0}$ at rest having the probability
$P_{i}$ in the spin state $|\psi_{i}\rangle$. Then the spin density
operator $\rho$ is given by Eq. (\ref{eq:density-operator}). The
angular distribution for the decay daughter can be written as 
\begin{eqnarray}
\frac{dN}{d\Omega} & = & \sum_{i}P_{i}\bigg|\langle K^{+},\pi^{-}|\mathscr{S}|\psi_{i}\rangle\bigg|^{2}\nonumber \\
 & = & \sum_{i}P_{i}\langle K^{+},\pi^{-}|\mathscr{S}|\psi_{i}\rangle\langle\psi_{i}|\mathscr{S}^{\dagger}|K^{+},\pi^{-}\rangle\nonumber \\
 & = & \langle K^{+},\pi^{-}|\mathscr{S}\rho\mathscr{S}^{\dagger}|K^{+},\pi^{-}\rangle.\label{eq:me5}
\end{eqnarray}
Inserting a completeness relation $\sum_{S_{z}}|K^{*0};S_{z}\rangle\langle K^{*0};S_{z}|=1$
into (\ref{eq:me5}), we obtain Eq. (10) of Ref. \cite{Schilling:1969um},

\begin{eqnarray}
\frac{dN}{d\Omega} & = & \sum_{S_{z1},S_{z2}}\langle K^{+},\pi^{-}|\mathscr{S}|K^{*0};S_{z1}\rangle\nonumber \\
 &  & \langle K^{*0};S_{z1}|\rho|K^{*0};S_{z2}\rangle\langle K^{*0};S_{z2}|\mathscr{S}^{\dagger}|K^{+},\pi^{-}\rangle\nonumber \\
 & = & \sum_{S_{z1},S_{z2}}\rho_{S_{z1},S_{z2}}Y_{1,S_{z1}}(\theta,\phi)Y_{1,S_{z2}}^{*}(\theta,\phi)\nonumber \\
 & = & \frac{3}{8\pi}\left[(1-\rho_{00})+(3\rho_{00}-1)\cos^{2}\theta\right.\nonumber \\
 &  & -2\text{Re}\rho_{-1,1}\sin^{2}\theta\cos(2\phi)-2\text{Im}\rho_{-1,1}\sin^{2}\theta\sin(2\phi)\nonumber \\
 &  & +\sqrt{2}\text{Re}(\rho_{-1,0}-\rho_{01})\sin(2\theta)\cos\phi\nonumber \\
 &  & \left.+\sqrt{2}\text{Im}(\rho_{-1,0}-\rho_{01})\sin(2\theta)\sin\phi\right],\label{eq:me6}
\end{eqnarray}
where $\rho_{S_{z1},S_{z2}}\equiv\langle K^{*0};S_{z1}|\rho|K^{*0};S_{z2}\rangle$
denote the spin density matrix elements for $K^{*0}$. Note that $dN/d\Omega$
is automatically normalized to 1, $\int d\Omega(dN/d\Omega)=1$. If
the parent particle is unpolarized, $\rho_{S_{z1},S_{z2}}=\delta_{S_{z1},S_{z2}}/3$,
then we have $dN/d\Omega=1/(4\pi)$.

In principle, if one measures $dN/d\Omega$ in experiments, one can
determine five real parameters out of eight ones in the spin density
matrix for vector mesons according to Eq. (\ref{eq:me6}), 
\begin{equation}
\rho_{00},\ \ \text{Re}\rho_{-1,1},\ \ \text{Im}\rho_{-1,1},\ \ \text{Re}(\rho_{-1,0}-\rho_{01}),\ \ \text{Im}(\rho_{-1,0}-\rho_{01}).\label{eq:you1}
\end{equation}
The polarization vector $\overrightarrow{\mathscr{P}}$ can be determined
from the spin density matrix \cite{Becattini:2016gvu} 
\begin{equation}
\overrightarrow{\mathscr{P}}=\frac{1}{S}\frac{\text{Tr}(\rho\hat{\mathbf{S}})}{\text{Tr}(\rho)},\label{eq:polar-density}
\end{equation}
where $S=1$ and the spin operators for vector mesons, $\hat{\mathbf{S}}=(\hat{S}_{1},\hat{S}_{2},\hat{S}_{3})$,
are defined by 
\begin{equation}
\hat{S}_{1}=\frac{1}{\sqrt{2}}\left(\begin{array}{ccc}
0 & 1 & 0\\
1 & 0 & 1\\
0 & 1 & 0
\end{array}\right),\ \ \hat{S}_{2}=\frac{1}{\sqrt{2}}\left(\begin{array}{ccc}
0 & -i & 0\\
i & 0 & -i\\
0 & i & 0
\end{array}\right),\ \ \hat{S}_{3}=\left(\begin{array}{ccc}
1 & 0 & 0\\
0 & 0 & 0\\
0 & 0 & -1
\end{array}\right).
\end{equation}
Then we obtain the polarization $\overrightarrow{\mathscr{P}}=(\mathscr{P}_{1},\mathscr{P}_{2},\mathscr{P}_{3})$,
\begin{eqnarray}
\mathscr{P}_{1} & = & \sqrt{2}\frac{1}{\text{Tr}(\rho)}\text{Re}(\rho_{-1,0}+\rho_{01}),\nonumber \\
\mathscr{P}_{2} & = & \sqrt{2}\frac{1}{\text{Tr}(\rho)}\text{Im}(\rho_{-1,0}+\rho_{01}),\nonumber \\
\mathscr{P}_{3} & = & \frac{1}{\text{Tr}(\rho)}\left(\rho_{11}-\rho_{-1,-1}\right).\label{eq:polar-vector}
\end{eqnarray}
Note that the above do not match the five quantities in (\ref{eq:you1})
that can be measured from $dN/d\Omega$. Therefore the polarization
of vector mesons cannot be measured from $dN/d\Omega$ in strong decays
in which the parity is conserved. This is very different from $\Lambda^{0}$
in the weak decay where the broken parity can be used to determine
the polarization.

When the azimuthal angle $\phi$ in Eq. (\ref{eq:me6}) is integrated
out, we obtain the polar angle distribution 
\begin{equation}
\frac{dN}{d\cos\theta}=\int_{0}^{2\pi}d\phi\frac{dN}{d\Omega}=\frac{3}{4}\left[(1-\rho_{00})+(3\rho_{00}-1)\cos^{2}\theta\right].\label{eq:density-m-00}
\end{equation}
So one can measure $\rho_{00}$ from the polar angle distribution
of the daughter particle. Any deviation from $1/3$ for $\rho_{00}$
may indicate some degree of polarization of vector mesons. This type
of measurements have been performed in the STAR experiment \cite{Abelev:2008ag}.

The third example is the spin-3/2 decuplet baryon $\Delta^{++}$ which
mostly decays into a proton and a $\pi^{+}$ by strong interaction.
The spin-parity of $\Delta^{++}$ is $(3/2)^{+}$, while those of
the proton and pion are $(1/2)^{+}$ and $0^{-}$ respectively. By
the parity conservation in the strong decay the angular distribution
of the proton must be in the P-wave state with $L=1$. The spin density
matrix of the spin-3/2 particle like $\Delta^{++}$ is a $4\times4$
complex matrix with fifteen independent parameters. The angular distribution
for the decay daughter can be written as 
\begin{eqnarray}
\frac{dN}{d\Omega} & = & \sum_{S_{z}^{\mathrm{p}}}\langle\hat{\mathbf{p}},S_{z}^{\mathrm{p}};\pi^{+}|\mathscr{S}\rho\mathscr{S}^{\dagger}|\hat{\mathbf{p}},S_{z}^{\mathrm{p}};\pi^{+}\rangle\nonumber \\
 & = & \sum_{S_{z1},S_{z2}}\rho_{S_{z1},S_{z2}}\sum_{S_{z}^{\mathrm{p}}}f(S_{z}^{\mathrm{p}},S_{z1})f^{*}(S_{z}^{\mathrm{p}},S_{z2}),\label{eq:angular-dist}
\end{eqnarray}
where the final state of the proton and pion is denoted as $\left|\hat{\mathbf{p}},S_{z}^{\mathrm{p}};\pi^{+}\right\rangle $
with $\hat{\mathbf{p}}$ being the proton's momentum direction and
$S_{z}^{\mathrm{p}}$ denoting the proton's spin state along the z-direction.
In deriving the second line of Eq. (\ref{eq:angular-dist}) we have
inserted into the first line of Eq. (\ref{eq:angular-dist}) a complete
set of spin states of $\Delta^{++}$, $\sum_{S_{z}}\left|\Delta^{++},S_{z}\right\rangle \left\langle \Delta^{++},S_{z}\right|=1$,
where $S_{z}$ denotes the spin state along the z-direction. In Eq.
(\ref{eq:angular-dist}), $f(S_{z}^{\mathrm{p}},S_{z})$ is the decay
amplitude given by Eq. (\ref{eq:decay-am}) and $\rho_{S_{z1},S_{z2}}$
are the spin density matrix elements for $\Delta^{++}$ given by Eq.
(\ref{eq:density-m}). The summation over $S_{z1}$ and $S_{z2}$
in Eq. (\ref{eq:angular-dist}) is made for all spin states of $\Delta^{++}$
with $S_{z1},S_{z2}=\pm\frac{1}{2},\pm\frac{3}{2}$. The final form
of the angular distribution in Eq. (\ref{eq:angular-dist}) is shown
in Eq. (\ref{eq:angular-dist-delta}). We see that the angular distribution
involves only five parameters out of fifteen independent ones, the
rest ten parameters cannot be determined by the daughter particle's
mometum distribution alone.

The polarization vector for $\Delta^{++}$ can be obtained from Eq.
(\ref{eq:polar-density}) and the result is given by Eq. (\ref{eq:polar-delta}).
We see that the parameters appearing in the polarization vector are
independent of those in the angular distribution (\ref{eq:angular-dist-delta}).
This is the feature of the parity conserved decay of $\Delta^{++}$,
same as in the strong decay of $K^{*0}$ in (\ref{eq:decay-vector}).

\section{Spin density matrix for vector mesons}

\label{sec:density-meson} We consider an ensemble of free quarks
and antiquarks in a deconfined system. These quarks and antiquarks
will combine to form hadrons in the hadronization process. For mesons
we define a quark-antiquark state with momenta and spins along a fixed
direction (the z-direction) 
\begin{equation}
\left|\mathrm{q}_{1},\bar{\mathrm{q}}_{2};s_{1},s_{2};\mathbf{p}_{1},\mathbf{p}_{2}\right\rangle \equiv\left|\mathrm{q}_{1},\bar{\mathrm{q}}_{2};s_{1},s_{2}\right\rangle \left|\mathrm{q};\mathbf{p}_{1},\mathbf{p}_{2}\right\rangle ,\label{eq:q-q-bar-st}
\end{equation}
where $\mathrm{q}_{1}=\mathrm{u},\mathrm{d},\mathrm{s}$ and $\bar{\mathrm{q}}_{2}=\bar{\mathrm{u}},\bar{\mathrm{d}},\bar{\mathrm{s}}$
denote the flavors of the quark and anti-quark respectively, $s_{1},s_{2}=\pm1/2$
denote the spin of the quark and antiquark in the z-direction respectively,
and 'q' labels the quark-antiquark momentum state. The momentum states
of the quark and antiquark in coordinate representation are plane
waves, 
\begin{equation}
\left\langle \mathbf{x}_{1},\mathbf{x}_{2}|\mathrm{q};\mathbf{p}_{1},\mathbf{p}_{2}\right\rangle =\frac{1}{V}\exp\left(i\mathbf{p}_{1}\cdot\mathbf{x}_{1}+i\mathbf{p}_{2}\cdot\mathbf{x}_{2}\right),\label{eq:quark-wave}
\end{equation}
where $V$ is the volume. Now we can write down the density operator
for quarks and antiquarks, 
\begin{eqnarray}
\rho & = & V^{2}\sum_{s_{1},s_{2}}\sum_{\mathrm{q}_{1},\bar{\mathrm{q}}_{2}}\int\frac{d^{3}\mathbf{p}_{1}}{(2\pi)^{3}}\frac{d^{3}\mathbf{p}_{2}}{(2\pi)^{3}}w_{\mathrm{q}1,s1}(\mathbf{p}_{1})w_{\bar{\mathrm{q}}2,s2}(\mathbf{p}_{2})\nonumber \\
 &  & \times\left|\mathrm{q}_{1},\bar{\mathrm{q}}_{2};s_{1},s_{2};\mathbf{p}_{1},\mathbf{p}_{2}\right\rangle \left\langle \mathrm{q}_{1},\bar{\mathrm{q}}_{2};s_{1},s_{2};\mathbf{p}_{1},\mathbf{p}_{2}\right|,
\end{eqnarray}
where $w_{\mathrm{q}1,s1}(\mathbf{p}_{1})$/$w_{\bar{\mathrm{q}}2,s2}(\mathbf{p}_{2})$
are the weight functions for the quark/antiquark which satisfy the
normalization conditions 
\begin{equation}
\sum_{s}w_{\mathrm{q},s}(\mathbf{p})=1,\quad\sum_{s}w_{\bar{\mathrm{q}},s}(\mathbf{p})=1.
\end{equation}
The weight functions are related to the polarization of the quark
and antiquark by 
\begin{eqnarray}
w_{\mathrm{q},\pm1/2}(\mathbf{p}) & = & \frac{1}{2}\left[1\pm\mathscr{P}_{\mathrm{q}}(\mathbf{p})\right],\nonumber \\
w_{\bar{\mathrm{q}},\pm1/2}(\mathbf{p}) & = & \frac{1}{2}\left[1\pm\mathscr{P}_{\bar{\mathrm{q}}}(\mathbf{p})\right],\label{eq:quark-weight}
\end{eqnarray}
where $\mathscr{P}_{\mathrm{q}}(\mathbf{p})$ and $\mathscr{P}_{\bar{\mathrm{q}}}(\mathbf{p})$
are the polarization of the quark and antiquark in the z-direction
respectively. We will use shorthand notations for the weight functions,
$w_{\mathrm{q}(\bar{\mathrm{q}}),\pm}\equiv w_{\mathrm{q}(\bar{\mathrm{q}}),\pm1/2}$,
in the rest of the paper. 

Now we look at the meson state defined by 
\begin{equation}
\left|\mathrm{M};S,S_{z},\mathbf{p}\right\rangle \equiv\left|\mathrm{M};S,S_{z}\right\rangle \left|\mathrm{M};\mathbf{p}\right\rangle ,
\end{equation}
where $\left|\mathrm{M};S,S_{z}\right\rangle $ denotes the meson's
spin-flavor wave function and 'M' labels the meson state. For a vector
meson we have $S=1$ and $S_{z}=-1,0,1$ for the spin state in the
z-direction. In the quark model, the momentum state of the meson in
coordinate representation is given by 
\begin{equation}
\left\langle \mathbf{x}_{1},\mathbf{x}_{2}|\mathrm{M};\mathbf{p}\right\rangle =\frac{1}{V^{1/2}}\exp(i\mathbf{p}\cdot\mathbf{x})\varphi_{\mathrm{M}}(\mathbf{y}).\label{eq:meson-wave}
\end{equation}
Here $\mathbf{x}=(\mathbf{x}_{1}+\mathbf{x}_{2})/2$ is the center
position of the quark and antiquark, $\mathbf{y}=\mathbf{x}_{1}-\mathbf{x}_{2}$
is their distance, and $\varphi_{\mathrm{M}}(\mathbf{y})$ is the
meson wave function satisfying the normailzation condition $\int d^{3}\mathbf{y}|\varphi_{\mathrm{M}}(\mathbf{y})|^{2}=1$.
We choose the Gaussian form for $\varphi_{\mathrm{M}}(\mathbf{y})$
whose Fourier transform is given by \cite{Greco:2003xt,Fries:2003kq}
\begin{eqnarray}
\varphi_{\mathrm{M}}(\mathbf{q}) & = & \int d^{3}\mathbf{y}e^{-i\mathbf{q}\cdot\mathbf{y}}\varphi_{\mathrm{M}}(\mathbf{y})\nonumber \\
 & = & \left(\frac{2\sqrt{\pi}}{a_{\mathrm{M}}}\right)^{3/2}\exp\left(-\frac{\mathbf{q}^{2}}{2a_{\mathrm{M}}^{2}}\right),\label{eq:meson-gauss}
\end{eqnarray}
where $a_{\mathrm{M}}$ is the coefficient characterizing the average
width of the momentum in the meson wave function. One can check that
$\varphi_{\mathrm{M}}(\mathbf{q})$ satisfies the normailzation condition
$\int d^{3}\mathbf{q}|\varphi_{\mathrm{M}}(\mathbf{q})|^{2}=(2\pi)^{3}$.

We can compute the spin density matrix element $\rho_{S_{z1}S_{z2}}$
on the vector meson state with, for example, the quark flavor $\mathrm{q}\bar{\mathrm{q}}$,
where $S_{z1},S_{z2}=-1,0,1$. The element $\rho_{00}$ for $S_{z1}=S_{z2}=0$
is of special interest since it can be measured in experiments (as
discussed in Section \ref{sec:density-matrix}), 
\begin{eqnarray}
\rho_{00}^{S=1}(\mathbf{p}) & = & \left\langle \mathrm{M};1,0,\mathbf{p}\right|\rho\left|\mathrm{M};1,0,\mathbf{p}\right\rangle \nonumber \\
 & = & \frac{1}{2}V^{2}\int\frac{d^{3}\mathbf{p}_{1}}{(2\pi)^{3}}\frac{d^{3}\mathbf{p}_{2}}{(2\pi)^{3}}\nonumber \\
 &  & \times\left[w_{\mathrm{q},+}(\mathbf{p}_{1})w_{\bar{\mathrm{q}},-}(\mathbf{p}_{2})+w_{\mathrm{q},-}(\mathbf{p}_{1})w_{\bar{\mathrm{q}},+}(\mathbf{p}_{2})\right]\left|\left\langle \mathrm{q};\mathbf{p}_{1},\mathbf{p}_{2}|\mathrm{M};\mathbf{p}\right\rangle \right|^{2}\nonumber \\
 & = & \frac{1}{2}\int\frac{d^{3}\mathbf{q}}{(2\pi)^{3}}\left[w_{\mathrm{q},+}\left(\frac{\mathbf{p}}{2}+\mathbf{q}\right)w_{\bar{\mathrm{q}},-}\left(\frac{\mathbf{p}}{2}-\mathbf{q}\right)\right.\nonumber \\
 &  & \left.+w_{\mathrm{q},-}\left(\frac{\mathbf{p}}{2}+\mathbf{q}\right)w_{\bar{\mathrm{q}},+}\left(\frac{\mathbf{p}}{2}-\mathbf{q}\right)\right]\left|\varphi_{\mathrm{M}}(\mathbf{q})\right|^{2},\label{eq:density-matrix-1}
\end{eqnarray}
where we have used the Clebsch-Gordan coefficients in the quark spin
wave function of the vector meson state $\left|S=1,S_{z}=0\right\rangle $
and the amplitude of a vector meson state projected onto the quark-antiquark
state 
\begin{eqnarray}
\left\langle \mathrm{q};\mathbf{p}_{1},\mathbf{p}_{2}|\mathrm{M};\mathbf{p}\right\rangle  & = & \int d^{3}\mathbf{x}_{1}d^{3}\mathbf{x}_{2}\left\langle \mathrm{q};\mathbf{p}_{1},\mathbf{p}_{2}|\mathbf{x}_{1},\mathbf{x}_{2}\right\rangle \left\langle \mathbf{x}_{1},\mathbf{x}_{2}|\mathrm{M};\mathbf{p}\right\rangle \nonumber \\
 & = & \frac{(2\pi)^{3}}{V^{3/2}}\delta^{(3)}(\mathbf{p}-\mathbf{p}_{1}-\mathbf{p}_{2})\varphi_{\mathrm{M}}(\mathbf{q}),\label{eq:amp-q-m}
\end{eqnarray}
where we have used Eqs. (\ref{eq:quark-wave},\ref{eq:meson-wave},\ref{eq:meson-gauss})
with $\mathbf{q}=(\mathbf{p}_{1}-\mathbf{p}_{2})/2$. From Eq. (\ref{eq:amp-q-m})
we obtain 
\begin{eqnarray}
\left|\left\langle \mathrm{q};\mathbf{p}_{1},\mathbf{p}_{2}|\mathrm{M};\mathbf{p}\right\rangle \right|^{2} & = & \frac{(2\pi)^{3}}{V^{2}}\delta^{(3)}(\mathbf{p}-\mathbf{p}_{1}-\mathbf{p}_{2})\left|\varphi_{\mathrm{M}}(\mathbf{q})\right|^{2},\label{eq:amp-qm-square}
\end{eqnarray}
where we have used $\delta^{(3)}(\mathbf{0})=V/(2\pi)^{3}$. Eq. (\ref{eq:amp-qm-square})
has been used in Eq. (\ref{eq:density-matrix-1}).

Similarly we can obtain other elements of the spin density matrix
for vector mesons as well as for scalar (spin-0) mesons. We list all
diagonal elements as follows, 
\begin{eqnarray}
\rho_{00}^{S=0}(\mathbf{p}) & = & \frac{1}{2}\int\frac{d^{3}\mathbf{q}}{(2\pi)^{3}}\left[w_{\mathrm{q},+}\left(\frac{\mathbf{p}}{2}+\mathbf{q}\right)w_{\bar{\mathrm{q}},-}\left(\frac{\mathbf{p}}{2}-\mathbf{q}\right)\right.\nonumber \\
 &  & \left.+w_{\mathrm{q},-}\left(\frac{\mathbf{p}}{2}+\mathbf{q}\right)w_{\bar{\mathrm{q}},+}\left(\frac{\mathbf{p}}{2}-\mathbf{q}\right)\right]\left|\varphi_{\mathrm{M}}(\mathbf{q})\right|^{2},\nonumber \\
\rho_{00}^{S=1}(\mathbf{p}) & = & \frac{1}{2}\int\frac{d^{3}\mathbf{q}}{(2\pi)^{3}}\left[w_{\mathrm{q},+}\left(\frac{\mathbf{p}}{2}+\mathbf{q}\right)w_{\bar{\mathrm{q}},-}\left(\frac{\mathbf{p}}{2}-\mathbf{q}\right)\right.\nonumber \\
 &  & \left.+w_{\mathrm{q},-}\left(\frac{\mathbf{p}}{2}+\mathbf{q}\right)w_{\bar{\mathrm{q}},+}\left(\frac{\mathbf{p}}{2}-\mathbf{q}\right)\right]\left|\varphi_{\mathrm{M}}(\mathbf{q})\right|^{2},\nonumber \\
\rho_{11}^{S=1}(\mathbf{p}) & = & \int\frac{d^{3}\mathbf{q}}{(2\pi)^{3}}w_{\mathrm{q},+}\left(\frac{\mathbf{p}}{2}+\mathbf{q}\right)w_{\bar{\mathrm{q}},+}\left(\frac{\mathbf{p}}{2}-\mathbf{q}\right)\left|\varphi_{\mathrm{M}}(\mathbf{q})\right|^{2},\nonumber \\
\rho_{-1,-1}^{S=1}(\mathbf{p}) & = & \int\frac{d^{3}\mathbf{q}}{(2\pi)^{3}}w_{\mathrm{q},-}\left(\frac{\mathbf{p}}{2}+\mathbf{q}\right)w_{\bar{\mathrm{q}},-}\left(\frac{\mathbf{p}}{2}-\mathbf{q}\right)\left|\varphi_{\mathrm{M}}(\mathbf{q})\right|^{2}.\label{eq:density-matrix}
\end{eqnarray}
Note that all off-diagonal elements for the spin triplet are vanishing,
i.e. $\rho_{S_{z1}S_{z2}}^{S=1}=0$ for $S_{z1}\neq S_{z2}$. Since
the trace of $\rho$ runs over all meson spin states, it fulfills
\begin{equation}
\rho_{00}^{S=0}(\mathbf{p})+\rho_{00}^{S=1}(\mathbf{p})+\rho_{11}^{S=1}(\mathbf{p})+\rho_{-1,-1}^{S=1}(\mathbf{p})=1.
\end{equation}
If we only consider vector mesons, we can obtain the normalized spin
density matrix element $\bar{\rho}_{S_{z}S_{z}}^{S=1}(\mathbf{p})$
($S_{z}=-1,0,1$) for the vector meson which is related to the quark
and antiquark polarization functions. The normalized spin density
matrix element $\bar{\rho}_{00}^{S=1}(\mathbf{p})$ is given by 
\begin{eqnarray}
\bar{\rho}_{00}^{S=1}(\mathbf{p}) & = & \frac{\int d^{3}\mathbf{q}[1-\mathscr{P}_{\mathrm{q}}(\mathbf{p}/2+\mathbf{q})\mathscr{P}_{\bar{\mathrm{q}}}(\mathbf{p}/2-\mathbf{q})]|\varphi_{\mathrm{M}}(\mathbf{q})|^{2}}{\int d^{3}\mathbf{q}[3+\mathscr{P}_{\mathrm{q}}(\mathbf{p}/2+\mathbf{q})\mathscr{P}_{\bar{\mathrm{q}}}(\mathbf{p}/2-\mathbf{q})]|\varphi_{\mathrm{M}}(\mathbf{q})|^{2}}.\label{eq:rho-00}
\end{eqnarray}
We see that if the quark and antiquark are not polarized ($\mathscr{P}_{\mathrm{q}}=\mathscr{P}_{\bar{\mathrm{q}}}=0$),
we have $\bar{\rho}_{00}^{S=1}=1/3$. If the quark polarization is
small, $\mathscr{P}_{\mathrm{q}}\sim\mathscr{P}_{\bar{\mathrm{q}}}\ll1$,
Eq. (\ref{eq:rho-00}) can be approximated as 
\begin{eqnarray}
\bar{\rho}_{00}^{S=1}(\mathbf{p}) & \approx & \frac{1}{3}-\frac{4}{9}\int\frac{d^{3}\mathbf{q}}{(2\pi)^{3}}\mathscr{P}_{\mathrm{q}}\left(\frac{\mathbf{p}}{2}+\mathbf{q}\right)\mathscr{P}_{\bar{\mathrm{q}}}\left(\frac{\mathbf{p}}{2}-\mathbf{q}\right)|\varphi_{\mathrm{M}}(\mathbf{q})|^{2}.\label{eq:rho00}
\end{eqnarray}
If the quark polarization is independent of momentum, Eq. (\ref{eq:rho00})
becomes 
\begin{equation}
\bar{\rho}_{00}^{S=1}\approx\frac{1}{3}-\frac{4}{9}\mathscr{P}_{\mathrm{q}}\mathscr{P}_{\bar{\mathrm{q}}}.\label{eq:rho-00-1}
\end{equation}
From Eqs. (\ref{eq:rho00},\ref{eq:rho-00-1}) we see that the deviation
from non-polarized case $\bar{\rho}_{00}^{S=1}=1/3$ is of second
order in $\mathscr{P}_{\mathrm{q}}$ or $\mathscr{P}_{\bar{\mathrm{q}}}$.

For $\phi(1020)$ with the quark content $\mathrm{s}\bar{\mathrm{s}}$,
we obtain its polarization along z-direction from Eq. (\ref{eq:polar-vector}),
\begin{equation}
\mathscr{P}_{\phi}=\frac{\rho_{11}^{\phi}-\rho_{-1-1}^{\phi}}{\rho_{11}^{\phi}+\rho_{00}^{\phi}+\rho_{-1-1}^{\phi}},\label{eq:polar-phi}
\end{equation}
where we have from Eq. (\ref{eq:density-matrix}), 
\begin{eqnarray*}
\rho_{11}^{\phi}-\rho_{-1,-1}^{\phi} & = & \frac{1}{2}\int\frac{d^{3}q}{(2\pi)^{3}}|\varphi_{\mathrm{M}}(\mathbf{q})|^{2}\left[\mathscr{P}_{\mathrm{s}}\left(\frac{\mathbf{p}}{2}+\mathbf{q}\right)+\mathscr{P}_{\bar{\mathrm{s}}}\left(\frac{\mathbf{p}}{2}-\mathbf{q}\right)\right],\\
\rho_{11}^{\phi}+\rho_{00}^{\phi}+\rho_{-1,-1}^{\phi} & = & \frac{3}{4}+\frac{1}{4}\int\frac{d^{3}q}{(2\pi)^{3}}|\varphi_{\mathrm{M}}(\mathbf{q})|^{2}\mathscr{P}_{\mathrm{s}}\left(\frac{\mathbf{p}}{2}+\mathbf{q}\right)\mathscr{P}_{\bar{\mathrm{s}}}\left(\frac{\mathbf{p}}{2}-\mathbf{q}\right).
\end{eqnarray*}
For the $\phi(1020)$ meson, $\rho_{00}^{\phi}$ is given by Eq. (\ref{eq:rho00})
or Eq. (\ref{eq:rho-00-1}) with replacement $\mathscr{P}_{\mathrm{q}}\rightarrow\mathscr{P}_{\mathrm{s}}$
and $\mathscr{P}_{\bar{\mathrm{q}}}\rightarrow\mathscr{P}_{\bar{\mathrm{s}}}$.
If the quark polarization is independent of momentum, Eq. (\ref{eq:polar-phi})
becomes 
\begin{eqnarray}
\mathscr{P}_{\phi} & = & \frac{2(\mathscr{P}_{\mathrm{s}}+\mathscr{P}_{\bar{\mathrm{s}}})}{3+\mathscr{P}_{\mathrm{s}}\mathscr{P}_{\bar{\mathrm{s}}}},\label{eq:phi-rho00}
\end{eqnarray}
From Section \ref{sec:density-matrix} we know that one cannot measure
the polarization $\mathscr{P}_{\phi}$ in the strong decay (\ref{eq:decay-vector})
except the element $\rho_{00}$, see Eq. (\ref{eq:density-m-00}).
However, by measuring $\rho_{00}^{\phi}$ one can determine the values
of $\mathscr{P}_{\mathrm{s}}$ and $\mathscr{P}_{\bar{\mathrm{s}}}$
from which one can indirectly determine $\mathscr{P}_{\phi}$.

\section{Spin density matrix for octet and decuplet baryons}

\label{sec:baryon}Similar to the quark-antiquark state defined in
Eqs. (\ref{eq:q-q-bar-st},\ref{eq:quark-wave}) for a meson, we can
define a three-quark state for a baryon in the deconfined quark system,
\begin{eqnarray}
\left|\mathrm{q}_{1},\mathrm{q}_{2},\mathrm{q}_{3};s_{1},s_{2},s_{3};\mathbf{p}_{1},\mathbf{p}_{2},\mathbf{p}_{3}\right\rangle  & \equiv & \left|\mathrm{q}_{1},s_{1},\mathbf{p}_{1}\right\rangle \left|\mathrm{q}_{2},s_{2},\mathbf{p}_{2}\right\rangle \left|\mathrm{q}_{3},s_{3},\mathbf{p}_{3}\right\rangle \nonumber \\
 & = & \left|\mathrm{q}_{1},\mathrm{q}_{2},\mathrm{q}_{3};s_{1},s_{2},s_{3}\right\rangle \left|\mathrm{q};\mathbf{p}_{1},\mathbf{p}_{2},\mathbf{p}_{3}\right\rangle ,\nonumber \\
\left\langle \mathbf{x}_{1},\mathbf{x}_{2},\mathbf{x}_{3}|\mathrm{q};\mathbf{p}_{1},\mathbf{p}_{2},\mathbf{p}_{3}\right\rangle  & = & \frac{1}{V^{3/2}}\exp\left(i\mathbf{p}_{1}\cdot\mathbf{x}_{1}+i\mathbf{p}_{2}\cdot\mathbf{x}_{2}+i\mathbf{p}_{3}\cdot\mathbf{x}_{3}\right),\label{eq:quark-f-s}
\end{eqnarray}
where $s_{1,2,3}=\pm1/2$ and $\mathrm{q}_{1,2,3}=\mathrm{u},\mathrm{d},\mathrm{s}$
denote the spin states in the z-direction and flavors of three quarks
respectively. The first line implies that the spin and flavor part
of the three-quark state is independent of its momentum part. The
second line means that three quarks are assumed to be in plane waves.
The spin density operator for three quarks has the form 
\begin{eqnarray}
\rho & = & V^{3}\sum_{s_{1},s_{2},s_{3}}\sum_{\mathrm{q}_{1},\mathrm{q}_{2},\mathrm{q}_{3}}\int\frac{d^{3}\mathbf{p}_{1}}{(2\pi)^{3}}\frac{d^{3}\mathbf{p}_{2}}{(2\pi)^{3}}\frac{d^{3}\mathbf{p}_{3}}{(2\pi)^{3}}w_{\mathrm{q}1,s1}(\mathbf{p}_{1})w_{\mathrm{q}2,s2}(\mathbf{p}_{2})w_{\mathrm{q}3,s3}(\mathbf{p}_{3})\nonumber \\
 &  & \times\left|\mathrm{q}_{1},\mathrm{q}_{2},\mathrm{q}_{3};s_{1},s_{2},s_{3};\mathbf{p}_{1},\mathbf{p}_{2},\mathbf{p}_{3}\right\rangle \left\langle \mathrm{q}_{1},\mathrm{q}_{2},\mathrm{q}_{3};s_{1},s_{2},s_{3};\mathbf{p}_{1},\mathbf{p}_{2},\mathbf{p}_{3}\right|,
\end{eqnarray}
where $w_{\mathrm{q},s}$ are the weight functions for quarks related
to the quark polarization $\mathscr{P}_{\mathrm{q}}$ by Eq. (\ref{eq:quark-weight}).

Now let us look at a baryon state defined by 
\begin{eqnarray}
\left|\mathrm{B};S,S_{z},\mathbf{p}\right\rangle  & = & \left|\mathrm{B};S,S_{z}\right\rangle \left|\mathrm{B};\mathbf{p}\right\rangle ,\label{eq:baryon-f-s}
\end{eqnarray}
where $\left|\mathrm{B};S,S_{z}\right\rangle $ is the spin-flavor
wave function of the baryon and $\left|\mathrm{B};\mathbf{p}\right\rangle $
is the baryon wave function in momentum space. For a spin-1/2 baryon,
we have $S=1/2$ and $S_{z}=\pm1/2$. The momentum state of the baryon
has the following form in coordinate representation 
\begin{eqnarray}
\left\langle \mathbf{x}_{1},\mathbf{x}_{2},\mathbf{x}_{3}|\mathrm{B};\mathbf{p}\right\rangle  & = & \frac{1}{V^{1/2}}\exp(i\mathbf{p}\cdot\mathbf{x})\varphi_{\mathrm{B}}(\mathbf{y},\mathbf{z}),\label{eq:baryon-wave}
\end{eqnarray}
where $\mathbf{x}=(\mathbf{x}_{1}+\mathbf{x}_{2}+\mathbf{x}_{3})/3$,
$\mathbf{y}=(\mathbf{x}_{1}+\mathbf{x}_{2})/2-\mathbf{x}_{3}$, and
$\mathbf{z}=\mathbf{x}_{1}-\mathbf{x}_{2}$ whose conjugate momenta
are $\mathbf{p}=\mathbf{p}_{1}+\mathbf{p}_{2}+\mathbf{p}_{3}$, $\mathbf{r}=(\mathbf{p}_{1}+\mathbf{p}_{2}-2\mathbf{p}_{3})/3$,
and $\mathbf{q}=(\mathbf{p}_{1}-\mathbf{p}_{2})/2$, respectively.
Note that the Jacobian between $(\mathbf{x}_{1},\mathbf{x}_{2},\mathbf{x}_{3})$
and $(\mathbf{x},\mathbf{y},\mathbf{z})$ is 1, so is the Jacobian
between $(\mathbf{p}_{1},\mathbf{p}_{2},\mathbf{p}_{3})$ and $(\mathbf{p},\mathbf{r},\mathbf{q})$.
The baryon wave function $\varphi_{\mathrm{B}}(\mathbf{y},\mathbf{z})$
is normalized as $\int d^{3}\mathbf{y}d^{3}\mathbf{z}\left|\varphi_{\mathrm{B}}(\mathbf{y},\mathbf{z})\right|^{2}=1$,
whose Fourier transform is assumed to have the Gaussian form \cite{Greco:2003xt,Fries:2003kq}
\begin{eqnarray}
\varphi_{\mathrm{B}}\left(\mathbf{r},\mathbf{q}\right) & = & \int d^{3}\mathbf{y}d^{3}\mathbf{z}\exp(-i\mathbf{q}\cdot\mathbf{z}-i\mathbf{r}\cdot\mathbf{y})\varphi_{\mathrm{B}}(\mathbf{y},\mathbf{z})\nonumber \\
 & = & (2\sqrt{\pi})^{3}\left(\frac{1}{a_{\mathrm{B}1}a_{\mathrm{B}2}}\right)^{3/2}\exp\left(-\frac{\mathbf{r}^{2}}{2a_{\mathrm{B}1}^{2}}-\frac{\mathbf{q}^{2}}{2a_{\mathrm{B}2}^{2}}\right),
\end{eqnarray}
where $a_{\mathrm{B}1}$ and $a_{\mathrm{B}2}$ are two coefficients
in the Gaussian wave packet for the baryon. One can check the normalization
condition $\int d^{3}\mathbf{r}d^{3}\mathbf{q}\left|\varphi_{\mathrm{B}}(\mathbf{r},\mathbf{q})\right|^{2}=(2\pi)^{6}$.

The spin density matrix elements for the baryon are given by projecting
the baryon states on the spin density operator, 
\begin{eqnarray}
\rho_{S_{z1}S_{z2}}(\mathbf{p}) & = & \left\langle \mathrm{B};S,S_{z1},\mathbf{p}\right|\rho\left|\mathrm{B};S,S_{z2},\mathbf{p}\right\rangle \nonumber \\
 & = & V^{3}\sum_{s_{1},s_{2},s_{3}}\sum_{\mathrm{q}_{1},\mathrm{q}_{2},\mathrm{q}_{3}}\int\frac{d^{3}\mathbf{p}_{1}}{(2\pi)^{3}}\frac{d^{3}\mathbf{p}_{2}}{(2\pi)^{3}}\frac{d^{3}\mathbf{p}_{3}}{(2\pi)^{3}}w_{\mathrm{q}1,s1}(\mathbf{p}_{1})w_{\mathrm{q}2,s2}(\mathbf{p}_{2})w_{\mathrm{q}3,s3}(\mathbf{p}_{3})\nonumber \\
 &  & \times\left\langle \mathrm{B};S_{z1}|\mathrm{q}_{1},\mathrm{q}_{2},\mathrm{q}_{3};s_{1},s_{2},s_{3}\right\rangle \left\langle \mathrm{q}_{1},\mathrm{q}_{2},\mathrm{q}_{3};s_{1},s_{2},s_{3}|\mathrm{B};S_{z2}\right\rangle \nonumber \\
 &  & \times\left|\left\langle \mathrm{B};\mathbf{p}|\mathrm{q};\mathbf{p}_{1},\mathbf{p}_{2},\mathbf{p}_{3}\right\rangle \right|^{2}.\label{eq:density-matrix-baryon}
\end{eqnarray}
Due to the factorization forms of the quark and baryon wave function
in Eqs. (\ref{eq:quark-f-s},\ref{eq:baryon-f-s}), we can calculate
the overlapping amplitudes of the spin-flavor parts and those of the
momentum parts separately.

We now look at the spin-flavor quark wave functions of spin-1/2 octet
baryons and spin-3/2 decuplet baryons which belong to the 56-plet
of SU(6) \cite{close}. The spin-flavor quark wave functions of these
baryons must be symmetric with repsect to any interchange of two quark
labels since their color wave function is anti-symmetric and their
orbital angular momentum is $L=0$. Therefore the SU(6) or spin-flavor
quark wave functions are in the form \cite{close} 
\begin{eqnarray}
\left|\mathrm{B};S=\frac{1}{2},S_{z}\right\rangle _{8} & = & \frac{1}{\sqrt{2}}\left(F_{\mathrm{MS}}\chi_{\mathrm{MS}}+F_{\mathrm{MA}}\chi_{\mathrm{MA}}\right)\nonumber \\
\left|\mathrm{B};S=\frac{3}{2},S_{z}\right\rangle _{10} & = & F_{\mathrm{S}}\chi_{\mathrm{S}}
\end{eqnarray}
where $F$ denotes the flavor wave function and $\chi$ denotes the
spin wave function with specific symmetry for interchange of two quark
labels, the supscript 8 and 10 mean octet and decuplet respectively,
and the subscript S, MS and MA mean symmetric, mixed symmetric and
mixed anti-symmetric respectively. 

Then we compute the spin density matrix elements for baryons in Eq.
(\ref{eq:density-matrix-baryon}). We take the $\Lambda$ hyperon
($\mathrm{B}=\Lambda$) as an example, whose flavor-spin parts of
the spin density matrix elements can be evaluated by the SU(6) quark
wave function of $\Lambda$. The projection amplitude of a baryon
momentum state on the three-quark one is given by 
\begin{eqnarray}
\left\langle \mathrm{q};\mathbf{p}_{1},\mathbf{p}_{2},\mathbf{p}_{3}|\mathrm{B};\mathbf{p}\right\rangle  & = & \int d^{3}\mathbf{x}_{1}d^{3}\mathbf{x}_{2}d^{3}\mathbf{x}_{3}\left\langle \mathrm{q};\mathbf{p}_{1},\mathbf{p}_{2},\mathbf{p}_{3}|\mathbf{x}_{1},\mathbf{x}_{2},\mathbf{x}_{3}\right\rangle \left\langle \mathbf{x}_{1},\mathbf{x}_{2},\mathbf{x}_{3}|\mathrm{B};\mathbf{p}\right\rangle \nonumber \\
 & = & \frac{1}{V^{2}}(2\pi)^{3}\delta^{(3)}(\mathbf{p}-\mathbf{p}_{1}-\mathbf{p}_{2}-\mathbf{p}_{3})\varphi_{\mathrm{B}}\left(\mathbf{r},\mathbf{q}\right),
\end{eqnarray}
whose modulus square is 
\begin{eqnarray}
\left|\left\langle \mathrm{q};\mathbf{p}_{1},\mathbf{p}_{2},\mathbf{p}_{3}|\mathrm{B};\mathbf{p}\right\rangle \right|^{2} & = & \frac{(2\pi)^{3}}{V^{3}}\delta^{(3)}(\mathbf{p}-\mathbf{p}_{1}-\mathbf{p}_{2}-\mathbf{p}_{3})\left|\varphi_{\mathrm{B}}\left(\mathbf{r},\mathbf{q}\right)\right|^{2}.\label{eq:baryon-momentum}
\end{eqnarray}
Then we obtain the diagonal element $\rho_{++}^{\Lambda}$ ($\equiv\rho_{\frac{1}{2},\frac{1}{2}}^{\Lambda}$)
in Eq. (\ref{eq:density-matrix-baryon}), 
\begin{eqnarray}
\rho_{++}^{\Lambda}(\mathbf{p}) & = & \left\langle \Lambda;\frac{1}{2},\frac{1}{2},\mathbf{p}\right|\rho\left|\Lambda;\frac{1}{2},\frac{1}{2},\mathbf{p}\right\rangle \nonumber \\
 & = & \frac{1}{24}\int\frac{d^{3}\mathbf{r}}{(2\pi)^{3}}\frac{d^{3}\mathbf{q}}{(2\pi)^{3}}\left|\varphi_{\Lambda}\left(\mathbf{r},\mathbf{q}\right)\right|^{2}\nonumber \\
 &  & \times\left\{ w_{\mathrm{s},+}(\mathbf{p}_{1})\left[2-\mathscr{P}_{\mathrm{u}}(\mathbf{p}_{2})\mathscr{P}_{\mathrm{d}}(\mathbf{p}_{3})-\mathscr{P}_{\mathrm{u}}(\mathbf{p}_{3})\mathscr{P}_{\mathrm{d}}(\mathbf{p}_{2})\right]\right.\nonumber \\
 &  & +w_{\mathrm{s},+}(\mathbf{p}_{2})\left[2-\mathscr{P}_{\mathrm{u}}(\mathbf{p}_{3})\mathscr{P}_{\mathrm{d}}(\mathbf{p}_{1})-\mathscr{P}_{\mathrm{u}}(\mathbf{p}_{1})\mathscr{P}_{\mathrm{d}}(\mathbf{p}_{3})\right]\nonumber \\
 &  & \left.+w_{\mathrm{s},+}(\mathbf{p}_{3})\left[2-\mathscr{P}_{\mathrm{u}}(\mathbf{p}_{1})\mathscr{P}_{\mathrm{d}}(\mathbf{p}_{2})-\mathscr{P}_{\mathrm{u}}(\mathbf{p}_{2})\mathscr{P}_{\mathrm{d}}(\mathbf{p}_{1})\right]\right\} ,\label{eq:rho++}
\end{eqnarray}
where we have applied $\mathbf{p}_{1}=\mathbf{p}/3+\mathbf{r}/2+\mathbf{q}$,
$\mathbf{p}_{2}=\mathbf{p}/3+\mathbf{r}/2-\mathbf{q}$ and $\mathbf{p}_{3}=\mathbf{p}/3-\mathbf{r}$.
Note that the subscript '$\mathrm{s},+$' of $w_{\mathrm{s},+}$ means
the s quark with spin up (please do not confuse s with the quark spin).
Similarly we obtain another diagonal element $\rho_{--}^{\Lambda}$
($\equiv\rho_{-\frac{1}{2},-\frac{1}{2}}^{\Lambda}$) 
\begin{eqnarray}
\rho_{--}^{\Lambda}(\mathbf{p}) & = & \left\langle \Lambda;\frac{1}{2},-\frac{1}{2},\mathbf{p}\right|\rho\left|\Lambda;\frac{1}{2},-\frac{1}{2},\mathbf{p}\right\rangle \nonumber \\
 & = & \frac{1}{24}\int\frac{d^{3}\mathbf{r}}{(2\pi)^{3}}\frac{d^{3}\mathbf{q}}{(2\pi)^{3}}\left|\varphi_{\Lambda}\left(\mathbf{r},\mathbf{q}\right)\right|^{2}\nonumber \\
 &  & \times\left\{ w_{\mathrm{s},-}(\mathbf{p}_{1})\left[2-\mathscr{P}_{\mathrm{u}}(\mathbf{p}_{2})\mathscr{P}_{\mathrm{d}}(\mathbf{p}_{3})-\mathscr{P}_{\mathrm{u}}(\mathbf{p}_{3})\mathscr{P}_{\mathrm{d}}(\mathbf{p}_{2})\right]\right.\nonumber \\
 &  & +w_{\mathrm{s},-}(\mathbf{p}_{2})\left[2-\mathscr{P}_{\mathrm{u}}(\mathbf{p}_{3})\mathscr{P}_{\mathrm{d}}(\mathbf{p}_{1})-\mathscr{P}_{\mathrm{u}}(\mathbf{p}_{1})\mathscr{P}_{\mathrm{d}}(\mathbf{p}_{3})\right]\nonumber \\
 &  & \left.+w_{\mathrm{s},-}(\mathbf{p}_{3})\left[2-\mathscr{P}_{\mathrm{u}}(\mathbf{p}_{1})\mathscr{P}_{\mathrm{d}}(\mathbf{p}_{2})-\mathscr{P}_{\mathrm{u}}(\mathbf{p}_{2})\mathscr{P}_{\mathrm{d}}(\mathbf{p}_{1})\right]\right\} .\label{eq:rpho--}
\end{eqnarray}
All off-diagonal elements are vanishing, i.e. $\rho_{+-}=\rho_{-+}=0$. 

The polarization formula for spin-1/2 baryons can be expressed as
\begin{eqnarray}
\mathscr{P}_{\mathrm{B,1/2}}(\mathbf{p}) & = & \frac{\rho_{++}^{\mathrm{B}}(\mathbf{p})-\rho_{--}^{\mathrm{B}}(\mathbf{p})}{\rho_{++}^{\mathrm{B}}(\mathbf{p})+\rho_{--}^{\mathrm{B}}(\mathbf{p})},\label{eq:polar-1/2-baryon}
\end{eqnarray}
where the superscript 'B' denotes a type of spin-1/2 baryons. For
the spin-3/2 baryons, the polarization is given by Eq. (\ref{eq:polar-delta}),
namely, 
\begin{equation}
\mathscr{P}_{\mathrm{B,3/2}}(\mathbf{p})=\frac{\frac{1}{3}\left[\rho_{\frac{1}{2},\frac{1}{2}}^{\mathrm{B}}(\mathbf{p})-\rho_{-\frac{1}{2},-\frac{1}{2}}^{\mathrm{B}}(\mathbf{p})\right]+\rho_{\frac{3}{2},\frac{3}{2}}^{\mathrm{B}}(\mathbf{p})-\rho_{-\frac{3}{2},-\frac{3}{2}}^{\mathrm{B}}(\mathbf{p})}{\rho_{\frac{1}{2},\frac{1}{2}}^{\mathrm{B}}(\mathbf{p})+\rho_{-\frac{1}{2},-\frac{1}{2}}^{\mathrm{B}}(\mathbf{p})+\rho_{\frac{3}{2},\frac{3}{2}}^{\mathrm{B}}(\mathbf{p})+\rho_{\frac{3}{2},\frac{3}{2}}^{\mathrm{B}}(\mathbf{p})},\label{eq:polar-3/2-baryon}
\end{equation}
where the superscript 'B' denotes a type of spin-3/2 baryons. 

For the $\Lambda$ hyperon we have 
\begin{eqnarray}
\rho_{++}^{\Lambda}(\mathbf{p})+\rho_{--}^{\Lambda}(\mathbf{p}) & = & \frac{1}{24}\int\frac{d^{3}\mathbf{r}}{(2\pi)^{3}}\frac{d^{3}\mathbf{q}}{(2\pi)^{3}}\left|\varphi_{\Lambda}\left(\mathbf{r},\mathbf{q}\right)\right|^{2}\nonumber \\
 &  & \times\left[6-\mathscr{P}_{\mathrm{u}}(\mathbf{p}_{2})\mathscr{P}_{\mathrm{d}}(\mathbf{p}_{3})-\mathscr{P}_{\mathrm{u}}(\mathbf{p}_{3})\mathscr{P}_{\mathrm{d}}(\mathbf{p}_{2})\right.\nonumber \\
 &  & -\mathscr{P}_{\mathrm{u}}(\mathbf{p}_{3})\mathscr{P}_{\mathrm{d}}(\mathbf{p}_{1})-\mathscr{P}_{\mathrm{u}}(\mathbf{p}_{1})\mathscr{P}_{\mathrm{d}}(\mathbf{p}_{3})\nonumber \\
 &  & \left.-\mathscr{P}_{\mathrm{u}}(\mathbf{p}_{1})\mathscr{P}_{\mathrm{d}}(\mathbf{p}_{2})-\mathscr{P}_{\mathrm{u}}(\mathbf{p}_{2})\mathscr{P}_{\mathrm{d}}(\mathbf{p}_{1})\right],\nonumber \\
\rho_{++}^{\Lambda}(\mathbf{p})-\rho_{--}^{\Lambda}(\mathbf{p}) & = & \frac{1}{24}\int\frac{d^{3}\mathbf{r}}{(2\pi)^{3}}\frac{d^{3}\mathbf{q}}{(2\pi)^{3}}\left|\varphi_{\Lambda}\left(\mathbf{r},\mathbf{q}\right)\right|^{2}\nonumber \\
 &  & \times\left\{ \mathscr{P}_{\mathrm{s}}(\mathbf{p}_{1})\left[2-\mathscr{P}_{\mathrm{u}}(\mathbf{p}_{2})\mathscr{P}_{\mathrm{d}}(\mathbf{p}_{3})-\mathscr{P}_{\mathrm{u}}(\mathbf{p}_{3})\mathscr{P}_{\mathrm{d}}(\mathbf{p}_{2})\right]\right.\nonumber \\
 &  & +\mathscr{P}_{\mathrm{s}}(\mathbf{p}_{2})\left[2-\mathscr{P}_{\mathrm{u}}(\mathbf{p}_{3})\mathscr{P}_{\mathrm{d}}(\mathbf{p}_{1})-\mathscr{P}_{\mathrm{u}}(\mathbf{p}_{1})\mathscr{P}_{\mathrm{d}}(\mathbf{p}_{3})\right]\nonumber \\
 &  & \left.+\mathscr{P}_{\mathrm{s}}(\mathbf{p}_{3})\left[2-\mathscr{P}_{\mathrm{u}}(\mathbf{p}_{1})\mathscr{P}_{\mathrm{d}}(\mathbf{p}_{2})-\mathscr{P}_{\mathrm{u}}(\mathbf{p}_{2})\mathscr{P}_{\mathrm{d}}(\mathbf{p}_{1})\right]\right\} .
\end{eqnarray}
Now we look at the limit that the quark polarization is much smaller
than 1. In this case we can neglect the quardratic terms in the quark
polarization, from Eq. (\ref{eq:polar-1/2-baryon}) the polarization
of $\Lambda$ can be approximated as 
\begin{equation}
\mathscr{P}_{\Lambda}(\mathbf{p})\approx\frac{1}{3}\int\frac{d^{3}\mathbf{r}}{(2\pi)^{3}}\frac{d^{3}\mathbf{q}}{(2\pi)^{3}}\left|\varphi_{\Lambda}\left(\mathbf{r},\mathbf{q}\right)\right|^{2}\left[\mathscr{P}_{\mathrm{s}}(\mathbf{p}_{1})+\mathscr{P}_{\mathrm{s}}(\mathbf{p}_{2})+\mathscr{P}_{\mathrm{s}}(\mathbf{p}_{3})\right],\label{eq:lambda-pol}
\end{equation}
which shows that the polarization of $\Lambda$ is mainly determined
by that of the s-quark. For antihyperons, we can make the replacement
$\Lambda\rightarrow\bar{\Lambda}$ and $\mathrm{u},\mathrm{d},\mathrm{s}\rightarrow\bar{\mathrm{u}},\bar{\mathrm{d}},\bar{\mathrm{s}}$
in Eqs. (\ref{eq:rho++}-\ref{eq:lambda-pol}). 

Similarly with the SU(6) quark wave functions of the octet baryons
$\Sigma^{0}$ and $\Xi^{-}$ and the decuplet baryons $\Delta^{++}$
and $\Omega$, we can obtain their polarizations from Eq. (\ref{eq:polar-1/2-baryon})
and (\ref{eq:polar-3/2-baryon}) respectively. 

If the quark polarization is independent of momentum, we obtain 
\begin{eqnarray}
\mathscr{P}_{\Lambda^{0}} & = & \mathscr{P}_{\mathrm{s}},\nonumber \\
\mathscr{P}_{\Sigma^{0}} & = & \frac{2(\mathscr{P}_{\mathrm{u}}+\mathscr{P}_{\mathrm{d}})-\mathscr{P}_{\mathrm{s}}-3\mathscr{P}_{\mathrm{u}}\mathscr{P}_{\mathrm{d}}\mathscr{P}_{\mathrm{s}}}{3-2\mathscr{P}_{\mathrm{s}}(\mathscr{P}_{\mathrm{u}}+\mathscr{P}_{\mathrm{d}})+\mathscr{P}_{\mathrm{u}}\mathscr{P}_{\mathrm{d}}},\nonumber \\
\mathscr{P}_{\Xi^{-}} & = & \frac{4\mathscr{P}_{\mathrm{s}}-\mathscr{P}_{\mathrm{d}}-3\mathscr{P}_{\mathrm{d}}\mathscr{P}_{\mathrm{s}}^{2}}{3-4\mathscr{P}_{\mathrm{d}}\mathscr{P}_{\mathrm{s}}+\mathscr{P}_{\mathrm{s}}^{2}},\nonumber \\
\mathscr{P}_{\Omega} & = & \frac{\mathscr{P}_{\mathrm{s}}(5+\mathscr{P}_{\mathrm{s}}^{2})}{3(1+\mathscr{P}_{\mathrm{s}}^{2})},\nonumber \\
\mathscr{P}_{\Delta^{++}} & = & \frac{\mathscr{P}_{\mathrm{u}}(5+\mathscr{P}_{\mathrm{u}}^{2})}{3(1+\mathscr{P}_{\mathrm{u}}^{2})}.\label{eq:polar-const-small}
\end{eqnarray}
If we further have $\mathscr{P}_{\mathrm{u}}=\mathscr{P}_{\mathrm{d}}\equiv\mathscr{P}_{\mathrm{q}}$,
the above polarization results are consistent to Ref. \cite{Liang:2004ph}.

\section{Fermion polarizations from Wigner functions}

\label{sec:wigner}The covariant Wigner function method \cite{Heinz:1983nx,Elze:1986qd,Vasak:1987um,Abada:1996hq}
for spin-1/2 fermions is a useful tool to study the chiral magnetic
effect, the chiral vortical effect and other related effects \cite{Gao:2012ix,Chen:2012ca,Gao:2015zka,Fang:2016vpj,Hidaka:2016yjf,Mueller:2017lzw}.
The Wigner function is equivalent to the quantum field and contains
all information that the quantum field does. One can also obtain the
quark polarization density from the Wigner function.

The Wigner function for spin-1/2 fermions in an external electromagnetic
field is given by \cite{Heinz:1983nx,Elze:1986qd,Vasak:1987um,Abada:1996hq}
\begin{equation}
W_{\alpha\beta}(x,p)=\frac{1}{(2\pi)^{4}}\int d^{4}ye^{-ip\cdot y}\left\langle \bar{\psi}_{\beta}\left(x+\frac{y}{2}\right)U(A;x+\frac{1}{2}y,x-\frac{1}{2}y)\psi_{\alpha}\left(x-\frac{y}{2}\right)\right\rangle ,
\end{equation}
where $\psi$ denotes the fermionic field and $\alpha,\beta$ the
spinor indices, $U(A;x_{2},x_{1})=\exp\left[iQ\int_{x_{1}}^{x_{2}}dx^{\mu}A_{\mu}(x)\right]$
is the gauge link which renders the Wigner function gauge invariant
with $A_{\mu}$ being the gauge potential of the electromagnetic field,
and $\left\langle \hat{O}\right\rangle $ denotes the ensemble average
of the operator $\hat{O}$ over thermal states. The spin tensor component
of the Wigner function can be extracted as 
\begin{eqnarray}
\mathscr{M}^{\mu\alpha\beta}(x,p) & \equiv & \frac{1}{2}\mathrm{Tr}\left[\{\gamma^{\mu},S^{\alpha\beta}\}W(x,p)\right]\nonumber \\
 & = & -\frac{1}{2}\epsilon^{\mu\alpha\beta\rho}\mathscr{A}_{\rho}(x,p),\label{eq:spin-tensor}
\end{eqnarray}
where $\mathscr{A}_{\rho}(x,p)$ is the axial vector component of
the Wigner function and can be obtained by $\mathscr{A}^{\mu}(x,p)=\text{Tr}[\gamma^{\mu}\gamma_{5}W(x,p)]$,
here we have used $\{\gamma^{\mu},S^{\nu\alpha}\}=\epsilon^{\mu\nu\alpha\lambda}\gamma^{5}\gamma_{\lambda}$
and $S^{\alpha\beta}=\frac{i}{4}[\gamma^{\alpha},\gamma^{\beta}]$.
Note that $\mathscr{M}^{\mu\alpha\beta}(x,p)$ are real quantities.
The spin angular momentum tensor of a fermionic system is given by
\begin{eqnarray}
M^{\rho\sigma} & = & \int d^{3}x\int d^{3}p\int dp_{0}u_{\lambda}\mathscr{M}^{\lambda\rho\sigma}(x,p)\nonumber \\
 & = & -\frac{1}{2}u_{\lambda}\epsilon^{\lambda\rho\sigma\alpha}\int d^{3}x\int d^{3}p\int dp_{0}\mathscr{A}_{\alpha}(x,p).\label{eq:m-a}
\end{eqnarray}
We see that $\int dp_{0}u_{\lambda}\mathscr{M}^{\lambda\rho\sigma}(x,p)$
plays the role of the spin tensor density in phase space. Here $u^{\lambda}$
is the fluid velocity and $p_{0}\equiv u\cdot p$. A four-momentum
can be decomposed as $p^{\lambda}=\bar{p}^{\lambda}+(p\cdot u)u^{\lambda}$.
So in a general frame the spatial momentum integral in Eq. (\ref{eq:m-a})
means $d^{3}p\equiv d^{3}\bar{p}$. Then an on-shell momentum satisfies
$p^{2}=m^{2}$ which can be rewritten as $p_{0}^{2}+\bar{p}^{2}=m^{2}$.
The on-shell energy is then given by $E_{p}\equiv\sqrt{m^{2}-\bar{p}^{2}}$.
In the local rest frame of the fluid we have $u^{\mu}=(1,\mathbf{0})$,
$\bar{p}^{\lambda}=(0,\mathbf{p})$, $\bar{p}^{2}=-|\mathbf{p}|^{2}$,
$E_{p}=\sqrt{|\mathbf{p}|^{2}+m^{2}}$. In this section we sometime
use the covariant form and sometime boldface version of a four-vector.
By boldface of the four-vector we refer to its the spatial component
in the local rest frame of the fluid.

The axial vector component at the zeroth or leading order in powers
of the field strength tensor and space-time derivative is given by
\cite{Fang:2016vpj} 
\begin{eqnarray}
\mathscr{A}_{(0)}^{\mu} & = & \mathrm{Tr}[\gamma^{\mu}\gamma^{5}W_{(0)}]\nonumber \\
 & = & m\left[\theta(p_{0})n^{\mu}(\mathbf{p},\mathbf{n})-\theta(-p_{0})n^{\mu}(-\mathbf{p},-\mathbf{n})\right]\delta(p^{2}-m^{2})(f_{+}-f_{-}),\label{eq:a0-1}
\end{eqnarray}
where $p_{0}\equiv u\cdot p$ and $f_{\pm}$ is given by 
\begin{equation}
f_{\pm}\equiv\frac{2}{(2\pi)^{3}}\left[\theta(p^{0})f_{\mathrm{FD}}(p_{0}-\mu_{\pm})+\theta(-p^{0})f_{\mathrm{FD}}(-p_{0}+\mu_{\pm})\right].\label{eq:fs}
\end{equation}
In Eqs. (\ref{eq:a0-1},\ref{eq:fs}) upper/lower sign denotes the
spin state along $\pm\mathbf{n}$ ($\mathbf{n}$ is the spin quantization
direction) in the particle's rest frame, $\mu_{\pm}$ denotes the
corresponding chemical potentials, and $f_{\mathrm{FD}}(E)=1/(e^{\beta E}+1)$
is the Fermi-Dirac distribution. In Eq. (\ref{eq:a0-1}) $n^{\mu}(\mathbf{p},\mathbf{n})$
is the spin quantization direction in the lab frame and given by 
\begin{eqnarray}
n^{\mu}(\mathbf{p},\mathbf{n}) & = & \Lambda_{\;\nu}^{\mu}(-\mathbf{v}_{p})n^{\nu}(\mathbf{0},\mathbf{n})=\left(\frac{\mathbf{n}\cdot\mathbf{p}}{m},\mathbf{n}+\frac{(\mathbf{n}\cdot\mathbf{p})\mathbf{p}}{m(m+E_{p})}\right).\label{eq:polar-cmoving}
\end{eqnarray}
Here $\Lambda_{\;\nu}^{\mu}(-\mathbf{v}_{p})$ is the Lorentz transformation
for $\mathbf{v}_{p}=\mathbf{p}/E_{p}$ and $n^{\nu}(\mathbf{0},\mathbf{n})=(0,\mathbf{n})$
is the four-vector of the spin quantization direction in the particle's
rest frame. One can check that $n^{\mu}(\mathbf{p},\mathbf{n})$ satisfies
$n^{2}=-1$ and $n\cdot p=0$, so it behaves like a spin four-vector
up to a factor of 1/2.

The first or next-to-leading order contribution for the axial vector
component of the Wigner function for massive fermions can be obtained
by generalizing the solution for massless fermions \cite{Gao:2012ix,Chen:2012ca,Gao:2015zka,Fang:2016vpj},
\begin{eqnarray}
\mathscr{A}_{(1)}^{\alpha}(x,p) & = & -\frac{1}{2}\hbar\beta\tilde{\Omega}^{\alpha\sigma}p_{\sigma}\delta(p^{2}-m^{2})\frac{d(f_{+}+f_{-})}{d(\beta p_{0})}-Q\hbar\tilde{F}^{\alpha\lambda}p_{\lambda}\frac{\delta(p^{2}-m^{2})}{p^{2}-m^{2}}(f_{+}+f_{-}),\label{eq:a1}
\end{eqnarray}
where $\tilde{F}^{\rho\lambda}=\frac{1}{2}\epsilon^{\rho\lambda\mu\nu}F_{\mu\nu}$
and $\tilde{\Omega}^{\xi\eta}=\frac{1}{2}\epsilon^{\xi\eta\nu\sigma}\Omega_{\nu\sigma}$
with $\Omega_{\nu\sigma}=\frac{1}{2}(\partial_{\nu}u_{\sigma}-\partial_{\sigma}u_{\nu})$.
Here $\epsilon^{\mu\nu\sigma\beta}$ and $\epsilon_{\mu\nu\sigma\beta}$
are anti-symmetric tensors with $\epsilon^{\mu\nu\sigma\beta}=1(-1)$
and $\epsilon_{\mu\nu\sigma\beta}=-1(1)$ for even (odd) permutations
of indices 0123, so we have $\epsilon^{0123}=-\epsilon_{0123}=1$.
In Eq. (\ref{eq:a1}) we have also assumed that $\beta=1/T$ is a
constant.

We now compute $\int dp_{0}A^{\rho}(x,p)$ from Eqs. (\ref{eq:a0-1},\ref{eq:a1}).
At the leading order, we obtain 
\begin{eqnarray}
\int dp_{0}A_{(0)}^{\alpha} & = & \frac{1}{(2\pi)^{3}}\frac{m}{E_{p}}\left[n^{\alpha}(\mathbf{p},\mathbf{n})\frac{1}{e^{\beta(E_{p}-\mu_{+})}+1}-n^{\alpha}(-\mathbf{p},-\mathbf{n})\frac{1}{e^{\beta(E_{p}+\mu_{+})}+1}\right.\nonumber \\
 &  & \left.-n^{\alpha}(\mathbf{p},\mathbf{n})\frac{1}{e^{\beta(E_{p}-\mu_{-})}+1}+n^{\alpha}(-\mathbf{p},-\mathbf{n})\frac{1}{e^{\beta(E_{p}+\mu_{-})}+1}\right].\label{eq:polar-1}
\end{eqnarray}
If $\mu_{\pm}=\mu$ does not depend on the spin state along $\pm\mathbf{n}$,
we see from Eq. (\ref{eq:polar-1}) that $A_{(0)}^{\alpha}=0$. In
this case the non-vanishing polarization can only come from the first-order
contribution $A_{(1)}^{\alpha}(x,p)$ in Eq. (\ref{eq:a1}) whose
vorticity part reads 
\begin{eqnarray}
\int dp_{0}A_{(\omega)}^{\alpha} & = & -\frac{1}{2}\hbar\beta\int dp_{0}\:\tilde{\Omega}^{\alpha\sigma}p_{\sigma}\frac{d(f_{+}+f_{-})}{d(\beta p_{0})}\delta(p^{2}-m^{2})\nonumber \\
 & = & \frac{1}{(2\pi)^{3}}\hbar\beta\frac{1}{E_{p}}\tilde{\Omega}^{\alpha\sigma}\left\{ \left.p_{\sigma}\right|_{p_{0}=E_{p}}\frac{e^{\beta(E_{p}-\mu)}}{[e^{\beta(E_{p}-\mu)}+1]^{2}}-\left.p_{\sigma}\right|_{p_{0}=-E_{p}}\frac{e^{\beta(E_{p}+\mu)}}{[e^{\beta(E_{p}+\mu)}+1]^{2}}\right\} \nonumber \\
 & \rightarrow & \frac{1}{(2\pi)^{3}}\hbar\beta\frac{1}{E_{p}}\tilde{\Omega}^{\alpha\sigma}p_{\sigma}\left\{ \frac{e^{\beta(E_{p}-\mu)}}{[e^{\beta(E_{p}-\mu)}+1]^{2}}+\frac{e^{\beta(E_{p}+\mu)}}{[e^{\beta(E_{p}+\mu)}+1]^{2}}\right\} ,\label{eq:spin-mass}
\end{eqnarray}
where in the last line we have made the replacement $\mathbf{p}\rightarrow-\mathbf{p}$
in the antifermion term and implied the same on-shell momentum $p^{\sigma}=(E_{p},\mathbf{p})$
for both fermions and antifermions. This makes no difference when
carrying out the integral over momentum. We have assumed the chemical
potentials are the same for both spin orientations, i.e. $\mu_{\pm}=\mu$,
so the sum over spin states gives a factor 2. Note that we can further
simplify Eq. (\ref{eq:spin-mass}) by keeping the time-like component
of the momentum, $p^{\sigma}\rightarrow u^{\sigma}E_{p}$, in the
last line if there is no dependence of $f_{\mathrm{FD}}(E_{p}\mp\mu)$
on the direction of $\mathbf{p}$. This will result in that the polarization
density is proportional to $\omega^{\alpha}=\tilde{\Omega}^{\alpha\sigma}u_{\sigma}$.
The electromagnetic field part $\int dp_{0}A_{(\mathrm{EM})}^{\alpha}$
from Eq. (\ref{eq:a1}) gives 
\begin{eqnarray}
\int dp_{0}A_{(\mathrm{EM})}^{\alpha} & = & -Q\hbar\int dp_{0}\:\tilde{F}^{\alpha\lambda}p_{\lambda}\frac{\delta(p^{2}-m^{2})}{p^{2}-m^{2}}(f_{+}+f_{-})\nonumber \\
 & \rightarrow & \frac{1}{(2\pi)^{3}}\beta Q\hbar\frac{1}{E_{p}^{2}}\tilde{F}^{\alpha\lambda}p_{\lambda}\left\{ \frac{e^{\beta(E_{p}-\mu)}}{[e^{\beta(E_{p}-\mu)}+1]^{2}}-\frac{e^{\beta(E_{p}+\mu)}}{[e^{\beta(E_{p}+\mu)}+1]^{2}}\right\} ,\label{eq:em-field-pi}
\end{eqnarray}
where we have used the same on-shell momentum $p^{\lambda}=(E_{p},\mathbf{p})$
for both fermions and antifermions. The derivation of (\ref{eq:em-field-pi})
is given in Appendix \ref{sec:derivation}.

We now obtain the spin tensor density in phase space, $\int dp_{0}u_{\lambda}\mathscr{M}^{\lambda\rho\sigma}(x,p)$,
in Eq. (\ref{eq:m-a}) as 
\begin{eqnarray}
\int dp_{0}u_{\lambda}\mathscr{M}^{\lambda\rho\sigma}(x,p) & = & -\frac{1}{2}u_{\lambda}\epsilon^{\lambda\rho\sigma\alpha}\int dp_{0}\mathscr{A}_{\alpha}(x,p),
\end{eqnarray}
where $\int dp_{0}\mathscr{A}_{\alpha}(x,p)$ are given by Eqs. (\ref{eq:spin-mass},\ref{eq:em-field-pi}).
It is convenient to distinguish fermions from antifermions, for which
we obtain their spin tensor densities in phase space 
\begin{eqnarray}
M_{\pm}^{\rho\sigma}(x,p) & = & -\frac{1}{2(2\pi)^{3}}\hbar\beta\frac{1}{E_{p}}u_{\lambda}\epsilon^{\lambda\rho\sigma\alpha}\left(\tilde{\Omega}_{\alpha\sigma}\pm Q\frac{1}{E_{p}}\tilde{F}_{\alpha\sigma}\right)p^{\sigma}\frac{e^{\beta(E_{p}\mp\mu)}}{[e^{\beta(E_{p}\mp\mu)}+1]^{2}},
\end{eqnarray}
where the upper/lower sign is for the fermion/antifermion. Using the
number density for fermions and antifermions in phase space, 
\begin{eqnarray}
n_{\pm}(x,p) & = & \frac{1}{(2\pi)^{3}}\frac{2}{e^{\beta(E_{p}\mp\mu)}+1},\label{eq:number-density}
\end{eqnarray}
we obtain the spin tensor per particle in phase space 
\begin{eqnarray}
\bar{M}_{\pm}^{\rho\sigma}(x,p) & = & \frac{M_{\pm}^{\rho\sigma}(x,p)}{n_{\pm}(x,p)}\nonumber \\
 & = & -\hbar\beta\frac{1}{4E_{p}}u_{\lambda}\epsilon^{\lambda\rho\sigma\alpha}\left(\tilde{\Omega}_{\alpha\sigma}\pm Q\frac{1}{E_{p}}\tilde{F}_{\alpha\sigma}\right)p^{\sigma}\left[1-f_{\text{FD}}(E_{p}\mp\mu)\right].
\end{eqnarray}
We then obtain the spin vector per particle in phase space for the
massive fermion or antifermion according to the Pauli-Lubanski pseudovector,
\begin{eqnarray}
S_{\pm}^{\mu}(x,p) & = & -\frac{1}{2m}\epsilon^{\mu\rho\sigma\nu}\bar{M}_{\rho\sigma}^{\pm}p_{\nu}\nonumber \\
 & = & \hbar\frac{\beta}{4m}\left(\tilde{\Omega}^{\mu\lambda}\pm Q\frac{1}{E_{p}}\tilde{F}^{\mu\lambda}\right)p_{\lambda}\left[1-f_{\text{FD}}(E_{p}\mp\mu)\right],
\end{eqnarray}
where $p$ is an on-shell four-momentum for both fermion and antifermion
and $E_{p}\equiv\sqrt{m^{2}-\bar{p}^{2}}=p\cdot u$ in a general frame.

In the non-relativistic limit and the local rest frame of the fluid,
the spatial component of the spin per particle in momentum space takes
the form 
\begin{eqnarray}
\mathbf{S}_{\pm}(\mathbf{p}) & \approx & \hbar\frac{\beta}{4m}\left(E_{p}\boldsymbol{\omega}\pm Q\mathbf{B}\right)\left[1-f_{\text{FD}}(E_{p}\mp\mu)\right],\label{polar-spin}
\end{eqnarray}
where we have dropped the coordinate dependence of the spin vector
for simplicity. The polarization per particle in momentum space can
be obtained by 
\begin{equation}
\overrightarrow{\mathscr{P}}_{\pm}(\mathbf{p})=2\mathbf{S}_{\pm}(\mathbf{p}).\label{eq:polar-momentum}
\end{equation}
We can apply Eqs. (\ref{polar-spin},\ref{eq:polar-momentum}) to
quarks and antiquarks and use Eq. (\ref{eq:rho00}) to compute $\rho_{00}$
for vector mesons and use Eq. (\ref{eq:lambda-pol}) to compute the
polarization of $\Lambda$ ($\bar{\Lambda}$).

\section{Polarizations: from quarks to vector mesons and baryons}

\label{sec:quark-polar} In this section we look closely at the quark
and hadron polarizations induced by the vorticity and magnetic field
by applying Eqs. (\ref{polar-spin},\ref{eq:polar-momentum}). We
look at the contributions from the vorticity and from the magnetic
field separately. For the simplicity of illustration we take a limit
with three conditions: (a) small constant polarizations; (b) Boltzmann
limit with $1-f_{\text{FD}}(E_{p}\mp\mu)\approx1$; (c) non-relativitic
limit with $E_{p}\approx m_{\mathrm{q}}$ ($m_{\mathrm{q}}$ is the
quark mass). At these limits the magnitudes of quark polarizations
have the simple form by Eqs. (\ref{polar-spin},\ref{eq:polar-momentum})
as 
\begin{eqnarray}
\mathscr{P}_{\mathrm{q}}(\omega) & \approx & \frac{1}{2}\beta\omega,\nonumber \\
\mathscr{P}_{\mathrm{q}}(B) & \approx & \beta\mu_{\mathrm{mq}}B,\label{eq:polar-vort-mag}
\end{eqnarray}
where $\mu_{\mathrm{mq}}=Q_{\mathrm{q}}/(2m_{\mathrm{q}})$ is the
magnetic moment of the quark with the electric charge $Q_{\mathrm{q}}$,
and we have used $\omega=|\boldsymbol{\omega}|$ and $B=|\mathbf{B}|$.
Substituting Eq. (\ref{eq:polar-vort-mag}) into Eq. (\ref{eq:polar-const-small}),
we obtain the polarizations of baryons to the leading order in the
quark polarization. We can also obtain the spin density matrix elements
and polarizations for the $\phi$ meson through Eqs. (\ref{eq:rho-00-1},\ref{eq:phi-rho00}).
All these results are listed in Table \ref{tab:polar}.

We make some remarks about these results. The first remark is about
the spin density matrix element of the $\phi$ meson $\rho_{00}^{\phi}$.
The deviations from the non-polarized value, $\rho_{00}^{\phi}-1/3$,
have the opposite sign from the vorticity and magnetic field: $\rho_{00}^{\phi}(\omega)-1/3\approx-(\beta\omega)^{2}/9<0$
while $\rho_{00}^{\phi}(B)-1/3\approx4\beta^{2}\mu_{\mathrm{ms}}^{2}B^{2}/9>0$.
Our result for $\rho_{00}^{\phi}(\omega)$ is consistent with the
hadron statitsical model \cite{Becattini:2016gvu} in which the mesons
are treated as elementary particles, but our result for $\rho_{00}^{\phi}(B)$
is different from Eqs. (\ref{eq:p-vector},\ref{eq:rho-00-vector})
in the hadron statitsical model \cite{Becattini:2016gvu}, see Appendix
\ref{sec:statisitical} for a brief introduction to the particle polarization
in the hadron statitsical model from the magnetic field and vorticity. 

In general cases, $\rho_{00}$ for other vector mesons is given by
Eq. (\ref{eq:rho-00-1}). Since the vorticity contribution $\rho_{00}(\omega)$
is independent of quark flavors, so we always have 
\begin{equation}
\rho_{00}(\omega)\approx\frac{1}{3}-\frac{1}{9}(\beta\omega)^{2}<\frac{1}{3},
\end{equation}
for all vector mesons. In contrast the magnetic field contribution
$\rho_{00}(B)$ does depend on the electric charges of the quark and
antiquark inside the vector meson, 
\begin{eqnarray}
\rho_{00}(B) & \approx & \frac{1}{3}-\frac{4}{9}\beta^{2}\mu_{\mathrm{m}\mathrm{q}_{1}}\mu_{\mathrm{m}\bar{\mathrm{q}}_{2}}B^{2}\nonumber \\
 & = & \frac{1}{3}-\frac{1}{9}\beta^{2}\frac{Q_{1}Q_{2}}{m_{1}m_{2}}B^{2},
\end{eqnarray}
where $Q_{1}$ and $Q_{2}$ are the electric charge of the quark and
antiquark in the vector meson respectively, and $m_{1}$ and $m_{2}$
are the mass of the quark and antiquark respectively. So for electrically
neutral vector mesons such as $\rho^{0}$, $K^{*0}$, $\bar{K}^{*0}$
and $\phi$, we have $\rho_{00}^{S=1}(B)>1/3$. But for electrically
charged vector mesons such as $\rho^{+}$, $\rho^{-}$, $K^{*+}$
and $K^{*-}$, we have $\rho_{00}^{S=1}(B)<1/3$. So we may conclude
that for electrically charged vector mesons we have $\rho_{00}<1/3$,
while the situation is inconclusive for electrically neutral vector
mesons, i.e. the magnitude of $\rho_{00}$ can be either $\rho_{00}<1/3$
or $\rho_{00}>1/3$ depending on the competition between the vorticity
and magnetic field contribution. 

The second remark is about the polarizations of baryons in magnetic
fields which we can express in terms of the baryon magnetic moments
with following formula in the constituent quark model \cite{close}:
\begin{eqnarray}
\mu_{\mathrm{m}\Lambda} & = & \mu_{\mathrm{ms}},\nonumber \\
\mu_{\mathrm{m}\Sigma} & = & \frac{1}{3}(2\mu_{\mathrm{mu}}+2\mu_{\mathrm{md}}-\mu_{\mathrm{ms}}),\nonumber \\
\mu_{\mathrm{m}\Xi} & = & \frac{1}{3}(4\mu_{\mathrm{ms}}-\mu_{\mathrm{md}}),\nonumber \\
\mu_{\mathrm{m}\Delta^{++}} & = & 3\mu_{\mathrm{mu}},\nonumber \\
\mu_{\mathrm{m}\Omega} & = & 3\mu_{\mathrm{ms}}.
\end{eqnarray}
By these relations in the constituent quark model about baryon magnetic
moments, our results for these baryons' polarizations are consistent
with the statistical model for hadrons \cite{Becattini:2016gvu} in
which hadrons are treated as elementary particles, see Appendix \ref{sec:statisitical}
for details. Our results for baryon polarizations from the vorticity
are also consistent with the hadron statistical model \cite{Becattini:2016gvu}.
It is also interesting to look at the ratios among the polarizations
of different baryons: 
\begin{eqnarray}
\mathscr{P}(\omega) & \rightarrow & \Lambda:\Sigma^{0}:\Xi^{-}:\Delta^{++}:\Omega=\frac{1}{2}:\frac{1}{2}:\frac{1}{2}:\frac{5}{6}:\frac{5}{6},\nonumber \\
\mathscr{P}(B) & \rightarrow & \Lambda:\Sigma^{0}:\Xi^{-}:w\Delta^{++}:\Omega=\mu_{\mathrm{m}\Lambda}:\mu_{\mathrm{m}\Sigma^{0}}:\mu_{\mathrm{m}\Xi^{-}}:\frac{5}{9}\mu_{\mathrm{m}\Delta^{++}}:\frac{5}{9}\mu_{\mathrm{m}\Omega},
\end{eqnarray}
where the first line is the ratios for the vorticity contributions
while the second line is for the magnetic field contributions. 

The final remark is that the polarization of any hadron from the vorticity
in the hadron statistical model is consistent with that in the quark
coalescence model under the three conditions as listed in the beginning
of this section. 

\begin{table}
\caption{\label{tab:polar}The polarizations of the $\phi$ meson and baryons
in the quark coalescence model. The spin density matrix elements $\rho_{00}^{\phi}$
for the $\phi$ meson are also listed. In the fisrt/second line are
listed the results for the $\phi$ meson and baryons from the vorticity/magnetic
field. In the third line are listed the polarizations of baryons in
terms of baryon magnetic moments given by the constituent quark model.
These results are consistent with those from the hadron statistical
model in which hadrons are treated as elementary particles. In the
third line are also listed the results for the $\phi$ meson in the
hadron statistical model in which $\phi$ is treated as an elementary
particle. The results are all expanded to the leading order in the
small polarization. }

\begin{tabular}{|c|c|c|c|c|c|c|c|}
\hline 
 & $\mathscr{P}_{\phi}$  & $\rho_{00}^{\phi}$  & $\mathscr{P}_{\Lambda}$  & $\mathscr{P}_{\Sigma^{0}}$  & $\mathscr{P}_{\Xi^{-}}$  & $\mathscr{P}_{\Delta^{++}}$  & $\mathscr{P}_{\Omega}$\tabularnewline
\hline 
$\mathscr{P}(\omega)$ or $\rho_{00}^{\phi}(\omega)$  & $\frac{2}{3}\beta\omega$  & $\frac{1}{3}-\frac{1}{9}(\beta\omega)^{2}$  & $\frac{1}{2}\beta\omega$  & $\frac{1}{2}\beta\omega$  & $\frac{1}{2}\beta\omega$  & $\frac{5}{6}\beta\omega$  & $\frac{5}{6}\beta\omega$\tabularnewline
\hline 
$\mathscr{P}(B)$ or $\rho_{00}^{\phi}(B)$  & $0$  & $\frac{1}{3}+\frac{4}{9}(\beta\mu_{\mathrm{ms}}B)^{2}$  & $\beta\mu_{\mathrm{ms}}B$  & $\frac{1}{3}\beta B(2\mu_{\mathrm{mu}}+2\mu_{\mathrm{md}}-\mu_{\mathrm{ms}})$  & $\frac{1}{3}\beta B(4\mu_{\mathrm{ms}}-\mu_{\mathrm{md}})$  & $\frac{5}{3}\beta\mu_{\mathrm{mu}}B$  & $\frac{5}{3}\beta\mu_{\mathrm{ms}}B$\tabularnewline
\hline 
$\mathscr{P}(B)$  & $\frac{2}{3}\beta\mu_{\mathrm{m}\phi}B$ & $\frac{1}{3}-\frac{1}{9}(\beta\mu_{\mathrm{m}\phi}B)^{2}$ & $\beta\mu_{\mathrm{m}\Lambda}B$  & $\beta\mu_{\mathrm{m}\Sigma^{0}}B$  & $\beta\mu_{\mathrm{m}\Xi^{-}}B$  & $\frac{5}{9}\beta\mu_{\mathrm{m}\Delta^{++}}B$  & $\frac{5}{9}\beta\mu_{\mathrm{m}\Omega}B$\tabularnewline
\hline 
\end{tabular}
\end{table}

\section{Summary and conclusions}

In this paper we formulate a non-relativistic quark coalesence model
with explicit momentum dependence based on the spin density matrix,
with which one can describe the spin alignments of vector mesons and
polarizations of baryons in a uniform way. The building blocks of
the quark coalescence model is the quark and antiquark polarizations
as the functions of momenta. The quark and antiquark polarizations
can be calculated from the spin-tensor component of the Wigner function.
Then we obtain the quark and antiquark polarizations induced by the
vorticity and the magnetic field, with which the polarizations of
vector mesons and baryons can be built up.

For vector mesons we start from the density matrix of quark-antiquark
states with propability functions related to the quark and antiquark
polarizations. Then we project vector meson states onto the density
matrix to obtain the spin density matrix element with momentum dependence.
The overlapping amplitude between the vector meson state and the quark-antiquark
state has to be evaluated. The final result for the spin density matrix
elements of vector mesons are obtained as a functional of the quark
and antiquark polarization functions. In the same way, we can also
describe the baryon polarizations in terms of the quark ones through
the density matrix of three-quark states. By projection of the baryon
state onto the density matrix of three-quark states we obtain the
spin density matrix of baryons as a functional of the quark and antiquark
polarization functions. The overlapping amplitude between the baryon
state and the three-quark state has to be evaluated. From the spin
density matrix of baryons we can compute the baryon polarizations. 

Note that the current quark coalescence model is non-relativistic
and valid for mesons and baryons with small momenta compared to their
masses. For hadrons with large momenta, one has to formulate a relativistic
version of the model. The current model can be the basis for further
numerical simulations with event generators to give realistic predictions
about global polarizations of vector mesons and baryons, which we
will investigate in the future.

\section*{Acknowledgments}

QW thanks H. Li and X.-L. Xia for helpful discussions. QW is supported
in part by the Major State Basic Research Development Program (973
program) in China under the Grant No. 2015CB856902 and 2014CB845402
and by the National Natural Science Foundation of China (NSFC) under
the Grant No. 11535012. 

\appendix

\section{Spin amplitudes and polarization vector of $\Delta^{++}$ }

In this appendix we give the spin amplitudes of the decay $\Delta^{++}\rightarrow\mathrm{p}+\pi^{+}$
and the polarization vector of $\Delta^{++}$. 

The initial spin states of $\Delta^{++}$ are written as $\left|\Delta^{++},S_{z}\right\rangle $,
where $S_{z}=\pm\frac{1}{2},\pm\frac{3}{2}$. The angular momentum
states of the proton and pion are in the multiplet of total angular
momentum $J=3/2$ which we denote as $\left|J,J_{z}\right\rangle _{\mathrm{f}}$.
The state $\left|J,J_{z}\right\rangle _{\mathrm{f}}$ is a coupled
state of the proton spin state $\left|\frac{1}{2},S_{z}^{\mathrm{p}}\right\rangle _{\mathrm{p}}$
and its orbital angular momentum state $\left|L,L_{z}\right\rangle _{\mathrm{L}}$
with $L=1$ and $L_{z}=0,\pm1$, 
\begin{eqnarray}
\left|\frac{3}{2},\frac{3}{2}\right\rangle _{\mathrm{f}} & = & \left|1,1\right\rangle _{\mathrm{L}}\left|\frac{1}{2},\frac{1}{2}\right\rangle _{\mathrm{p}},\nonumber \\
\left|\frac{3}{2},-\frac{3}{2}\right\rangle _{\mathrm{f}} & = & \left|1,-1\right\rangle _{\mathrm{L}}\left|\frac{1}{2},-\frac{1}{2}\right\rangle _{\mathrm{p}},\nonumber \\
\left|\frac{3}{2},\frac{1}{2}\right\rangle _{\mathrm{f}} & = & \sqrt{\frac{1}{3}}\left|1,1\right\rangle _{\mathrm{L}}\left|\frac{1}{2},-\frac{1}{2}\right\rangle _{\mathrm{p}}+\sqrt{\frac{2}{3}}\left|1,0\right\rangle _{\mathrm{L}}\left|\frac{1}{2},\frac{1}{2}\right\rangle _{\mathrm{p}},\nonumber \\
\left|\frac{3}{2},-\frac{1}{2}\right\rangle _{\mathrm{f}} & = & \sqrt{\frac{2}{3}}\left|1,0\right\rangle _{\mathrm{L}}\left|\frac{1}{2},-\frac{1}{2}\right\rangle _{\mathrm{p}}+\sqrt{\frac{1}{3}}\left|1,-1\right\rangle _{\mathrm{L}}\left|\frac{1}{2},\frac{1}{2}\right\rangle _{\mathrm{p}}.
\end{eqnarray}
We can define the decay transition matrix $\mathscr{S}$ as 
\begin{equation}
\mathscr{S}\left|\Delta^{++},S_{z}\right\rangle =\left|\frac{3}{2},S_{z}\right\rangle _{\mathrm{f}},
\end{equation}
from which we can compute the decay transition amplitude between the
initial and final state, 
\begin{eqnarray}
f(S_{z}^{\mathrm{p}},S_{z}) & = & \langle\hat{\mathbf{p}},S_{z}^{\mathrm{p}};\pi^{+}|\mathscr{S}|\Delta^{++},S_{z}\rangle\nonumber \\
 & = & \langle\hat{\mathbf{p}},S_{z}^{\mathrm{p}};\pi^{+}\left|\frac{3}{2},S_{z}\right\rangle _{\mathrm{f}}=\langle\hat{\mathbf{p}},S_{z}^{\mathrm{p}}\left|\frac{3}{2},S_{z}\right\rangle _{\mathrm{f}},\label{eq:decay-am}
\end{eqnarray}
where the final state can be denoted as $\left|\hat{\mathbf{p}},S_{z}^{\mathrm{p}};\pi^{+}\right\rangle \equiv\left|\theta,\phi,S_{z}^{\mathrm{p}}\right\rangle $
with $\hat{\mathbf{p}}=(\theta,\phi)$ being the proton's momentum
direction in the polar and azimuthal angle. With $\langle\hat{\mathbf{p}}|L=1,L_{z}\rangle_{\mathrm{L}}=Y_{1,L_{z}}(\theta,\phi)$,
we obtain $f(S_{z}^{\mathrm{p}},S_{z})$ as 
\begin{eqnarray}
f\left(\frac{1}{2},\frac{3}{2}\right) & = & Y_{1,1}(\theta,\phi),\;\; f\left(-\frac{1}{2},\frac{3}{2}\right)=0,\nonumber \\
f\left(-\frac{1}{2},-\frac{3}{2}\right) & = & Y_{1,-1}(\theta,\phi),\;\; f\left(\frac{1}{2},-\frac{3}{2}\right)=0,\nonumber \\
f\left(\frac{1}{2},\frac{1}{2}\right) & = & \sqrt{\frac{2}{3}}Y_{1,0}(\theta,\phi),\;\; f\left(-\frac{1}{2},\frac{1}{2}\right)=\sqrt{\frac{1}{3}}Y_{1,1}(\theta,\phi),\nonumber \\
f\left(\frac{1}{2},-\frac{1}{2}\right) & = & \sqrt{\frac{1}{3}}Y_{1,-1}(\theta,\phi),\;\; f\left(-\frac{1}{2},-\frac{1}{2}\right)=\sqrt{\frac{2}{3}}Y_{1,0}(\theta,\phi).\label{eq:decay-ampl}
\end{eqnarray}
With the decay amplitudes (\ref{eq:decay-ampl}), we can evaluate
Eq. (\ref{eq:angular-dist}) and obtain the angular distribution as
\begin{eqnarray}
\frac{dN}{d\Omega} & = & \frac{3}{8\pi}\bigg\{\bigg[1-\frac{2}{3}\bigg(\rho_{-\frac{1}{2},-\frac{1}{2}}+\rho_{\frac{1}{2}\frac{1}{2}}\bigg)\bigg]-\bigg[1-2\bigg(\rho_{-\frac{1}{2},-\frac{1}{2}}+\rho_{\frac{1}{2},\frac{1}{2}}\bigg)\bigg]\cos^{2}\theta\nonumber \\
 &  & +\frac{2}{\sqrt{3}}\bigg(\text{Re}\rho_{-\frac{3}{2},-\frac{1}{2}}-\text{Re}\rho_{\frac{1}{2},\frac{3}{2}}\bigg)\sin(2\theta)\cos\phi\nonumber \\
 &  & +\frac{2}{\sqrt{3}}\bigg(\text{Im}\rho_{-\frac{3}{2},-\frac{1}{2}}-\text{Im}\rho_{\frac{1}{2},\frac{3}{2}}\bigg)\sin(2\theta)\sin\phi\nonumber \\
 &  & -\frac{2}{\sqrt{3}}\bigg(\text{Re}\rho_{-\frac{3}{2},\frac{1}{2}}+\text{Re}\rho_{-\frac{1}{2},\frac{3}{2}}\bigg)\sin^{2}\theta\cos(2\phi)\nonumber \\
 &  & -\frac{2}{\sqrt{3}}\bigg(\text{Im}\rho_{-\frac{3}{2},\frac{1}{2}}+\text{Im}\rho_{-\frac{1}{2},\frac{3}{2}}\bigg)\sin^{2}\theta\sin(2\phi)\bigg\},\label{eq:angular-dist-delta}
\end{eqnarray}
where the spin density matrix elements for $\Delta^{++}$ are defined
by 
\begin{equation}
\rho_{S_{z1},S_{z2}}\equiv\langle\Delta^{++},S_{z1}|\rho|\Delta^{++},S_{z2}\rangle.\label{eq:density-m}
\end{equation}

The polarization vector $\overrightarrow{\mathscr{P}}=(\mathscr{P}_{1},\mathscr{P}_{2},\mathscr{P}_{3})$
for $\Delta^{++}$ can be determined from Eq. (\ref{eq:polar-density})
with $S=3/2$ and the spin operators for spin-3/2 particles being
defined by 
\begin{equation}
\hat{S}_{1}=\frac{1}{2}\left(\begin{array}{cccc}
0 & \sqrt{3} & 0 & 0\\
\sqrt{3} & 0 & 2 & 0\\
0 & 2 & 0 & \sqrt{3}\\
0 & 0 & \sqrt{3} & 0
\end{array}\right),\ \ \hat{S}_{2}=\frac{i}{2}\left(\begin{array}{cccc}
0 & -\sqrt{3} & 0 & 0\\
\sqrt{3} & 0 & -2 & 0\\
0 & 2 & 0 & -\sqrt{3}\\
0 & 0 & \sqrt{3} & 0
\end{array}\right),\ \ \hat{S}_{3}=\left(\begin{array}{cccc}
\frac{3}{2} & 0 & 0 & 0\\
0 & \frac{1}{2} & 0 & 0\\
0 & 0 & -\frac{1}{2} & 0\\
0 & 0 & 0 & -\frac{3}{2}
\end{array}\right),
\end{equation}
whose result is 
\begin{eqnarray}
\mathscr{P}_{1} & = & \frac{1}{\text{Tr}(\rho)}\left[\frac{2}{\sqrt{3}}\bigg(\text{Re}\rho_{-\frac{3}{2},-\frac{1}{2}}+\text{Re}\rho_{\frac{1}{2},\frac{3}{2}}\bigg)+\frac{4}{3}\text{Re}\rho_{-\frac{1}{2},\frac{1}{2}}\right],\nonumber \\
\mathscr{P}_{2} & = & \frac{1}{\text{Tr}(\rho)}\left[\frac{2}{\sqrt{3}}\bigg(\text{Im}\rho_{-\frac{3}{2},-\frac{1}{2}}+\text{Im}\rho_{\frac{1}{2},\frac{3}{2}}\bigg)+\frac{4}{3}\text{Im}\rho_{-\frac{1}{2},\frac{1}{2}}\right],\nonumber \\
\mathscr{P}_{3} & = & \frac{1}{\text{Tr}(\rho)}\left[\frac{1}{3}\bigg(\rho_{\frac{1}{2},\frac{1}{2}}-\rho_{-\frac{1}{2},-\frac{1}{2}}\bigg)+\rho_{\frac{3}{2},\frac{3}{2}}-\rho_{-\frac{3}{2},-\frac{3}{2}}\right].\label{eq:polar-delta}
\end{eqnarray}

\section{Derivation of Eq. (\ref{eq:em-field-pi})}

\label{sec:derivation}We give a detailed derivation of Eq. (\ref{eq:em-field-pi}).
To this end we look at the four-momentum integral, 

\begin{eqnarray}
\int d^{4}p\mathscr{A}_{(\mathrm{EM})}^{\mu} & = & Q\hbar\int d^{4}p\tilde{F}^{\mu\nu}p_{\nu}\delta^{\prime}(p^{2}-m^{2})(f_{+}+f_{-})\nonumber \\
 & = & \frac{4}{(2\pi)^{3}}Q\hbar\tilde{F}^{\mu\nu}\int d^{4}pp_{\nu}\delta^{\prime}(p^{2}-m^{2})\nonumber \\
 &  & \times\left[\theta(u\cdot p)f_{\mathrm{FD}}(u\cdot p-\mu)+\theta(-u\cdot p)f_{\mathrm{FD}}(-u\cdot p+\mu)\right]\nonumber \\
 & = & \frac{4}{(2\pi)^{3}}Q\hbar\tilde{F}^{\mu\nu}\int d^{4}pp_{\nu}\delta^{\prime}(p^{2}-m^{2})\theta(u\cdot p)\left[f_{\mathrm{FD}}(u\cdot p-\mu)-f_{\mathrm{FD}}(u\cdot p+\mu)\right]\nonumber \\
 & = & \frac{2}{(2\pi)^{3}}Q\hbar\tilde{F}^{\mu\nu}\int d^{4}p\frac{d\delta(p^{2}-m^{2})}{dp^{\nu}}\theta(u\cdot p)\left[f_{\mathrm{FD}}(u\cdot p-\mu)-f_{\mathrm{FD}}(u\cdot p+\mu)\right]\nonumber \\
 & = & -\frac{2}{(2\pi)^{3}}Q\hbar\tilde{F}^{\mu\nu}\int d^{4}p\delta(p^{2}-m^{2})\theta(u\cdot p)\frac{d}{dp^{\nu}}\left[f_{\mathrm{FD}}(u\cdot p-\mu)-f_{\mathrm{FD}}(u\cdot p+\mu)\right]\nonumber \\
 & = & \frac{2}{(2\pi)^{3}}\hbar Q\beta\tilde{F}^{\mu\nu}u_{\nu}\int d^{4}p\delta(p^{2}-m^{2})\theta(u\cdot p)\nonumber \\
 &  & \times\left[f_{\mathrm{FD}}(u\cdot p-\mu)(1-f_{\mathrm{FD}}(u\cdot p-\mu))-f_{\mathrm{FD}}(u\cdot p+\mu)(1-f_{\mathrm{FD}}(u\cdot p+\mu))\right]\nonumber \\
 & = & \frac{1}{(2\pi)^{3}}\hbar Q\beta\tilde{F}^{\mu\nu}\int d^{3}p\frac{p_{\nu}}{E_{p}^{2}}\nonumber \\
 &  & \times\left[f_{\mathrm{FD}}(E_{p}-\mu)(1-f_{\mathrm{FD}}(E_{p}-\mu))-f_{\mathrm{FD}}(E_{p}+\mu)(1-f_{\mathrm{FD}}(E_{p}+\mu))\right].
\end{eqnarray}
In the second line we have used $\mu_{\pm}=\mu$, then the sum over
spin states gives a factor 2. In the fifth line we have used the integral
by part and dropped the complete derivative term. In the last equality
we have replaced $u_{\nu}\rightarrow p_{\nu}/E_{p}$ and carried out
the integral over $p_{0}\equiv p\cdot u$, where $E_{p}=\sqrt{m^{2}-\bar{p}^{2}}$.
Then we arrive at Eq. (\ref{eq:em-field-pi}).

\section{Particle polarization in statistical model for hadrons}

\label{sec:statisitical}In the statisitical model for hadrons, we
treat a hadron as an elementary particle with inner structure. We
take the nonrelativistic limit and consider a hadron with the spin
$S$ and the magnetic moment $\boldsymbol{\mu}_{\mathrm{m}}=\mu_{\mathrm{m}}\mathbf{S}/S$
in a magnetic field $\mathbf{B}=B\mathbf{e}_{z}$ along the $z$-direction.
We know that the spin along the $z$-direction takes the value $S_{z}=-S,-S+1,\cdots,S-1,S$.
Suppose that the ensemble of the particles is in an equilibrium state
with the temperature $T$, so the probability for the state with a
specific value of $S_{z}$ is 
\begin{equation}
w(S_{z})=\frac{e^{\beta\mu_{\mathrm{m}}BS_{z}/S}}{\sum_{S_{z}^{\prime}=-S}^{S}e^{\beta\mu_{\mathrm{m}}BS_{z}^{\prime}/S}}.
\end{equation}
Then the average polarization along the $z$-direction is 
\begin{eqnarray}
\mathscr{P}_{S} & = & \frac{1}{S}\sum_{S_{z}=-S}^{S}S_{z}w\left(S_{z}\right)\nonumber \\
 & = & \left(1+\frac{1}{2S}\right)\coth\left[\left(1+\frac{1}{2S}\right)\beta\mu_{\mathrm{m}}B\right]-\frac{1}{2S}\coth\left(\frac{\beta\mu_{\mathrm{m}}B}{2S}\right).
\end{eqnarray}
If the magnetic field is weak, in the leading order we have 
\begin{equation}
\mathscr{P}_{S}\approx\frac{(S+1)}{3S}\beta\mu_{\mathrm{m}}B.
\end{equation}
For spin-1/2 and 3/2 particles, the above becomes 
\begin{equation}
\mathscr{P}_{S=1/2}=\beta\mu_{\mathrm{m},1/2}B,\ \ \mathscr{P}_{S=3/2}=\frac{5}{9}\beta\mu_{\mathrm{m},3/2}B,
\end{equation}
which are consistent with the quark coalescence model. 

For spin-1 particles, we have 
\begin{equation}
\mathscr{P}_{S=1}=\frac{2}{3}\beta\mu_{\mathrm{m},1}B.\label{eq:p-vector}
\end{equation}
We can also obtain the density matrix element in the weak magnetic
field 
\begin{eqnarray}
\rho_{00}^{S=1} & = & w(S=1,S_{z}=0)=\frac{1}{1+2\cosh(\beta\mu_{\mathrm{m},1}B)}\nonumber \\
 & \approx & \frac{1}{3}-\frac{1}{9}(\beta\mu_{\mathrm{m},1}B)^{2}\leq\frac{1}{3},\label{eq:rho-00-vector}
\end{eqnarray}
where the equality holds for particles without the magnetic moment.
For the $\phi$ meson, we have $\mathscr{P}_{\phi}=\frac{2}{3}\beta\mu_{\mathrm{m}\phi}B$
and $\rho_{\phi}=\frac{1}{3}-\frac{1}{9}(\beta\mu_{\mathrm{m}\phi}B)^{2}$. 

The polarization from the vorticity can be obatined by replacing $\mu_{\mathrm{m}}B/S$
by $\omega$. 

 \bibliographystyle{apsrev}
\bibliography{ref-1}

\end{document}